%% file: noebauer_sim.tex
\documentclass[usenatbib,useAMS]{mn2e}
\usepackage[T1]{fontenc}
\usepackage{aecompl}
\usepackage{graphicx}
\usepackage{times}
\usepackage{amssymb}
\usepackage{amsmath}
\usepackage{xspace}
\usepackage{siunitx}
\usepackage{tikz}
\usepackage{hyperref}

\usetikzlibrary{external, positioning, calc, through, shadows, fadings, patterns, mindmap, trees, arrows, shapes, backgrounds, decorations.markings, decorations.pathmorphing, decorations.pathreplacing, calc, 3d, intersections, shapes.gates.logic.US, external}

\newcommand{\athena}{\textsc{Athena}\xspace}
\newcommand{\python}{\textsc{Python}\xspace}
\newcommand{\tardis}{\textsc{Tardis}\xspace}
\newcommand{\mcrh}{\textsc{Mcrh}\xspace}

\newcommand{\cak}{\textsc{Cak}\xspace}
\newcommand{\mcak}{\textsc{Mcak}\xspace}
\newcommand{\mcakrh}{\textsc{Mcak-Rh}\xspace}
\newcommand{\cakrh}{\textsc{Cak-Rh}\xspace}

\def\araa{ARA\&A}%
\def\apj{ApJ}%
\def\apjs{ApJS}%
\def\apss{Ap\&SS}%
\def\aap{A\&A}%
\def\mnras{MNRAS}%
\title[Line-driven winds with Monte Carlo radiation hydrodynamics]{Self-consistent modelling of
line-driven hot-star winds with Monte Carlo radiation hydrodynamics}
\author[Noebauer~et~al.]{U.~M.~Noebauer,$^1$\thanks{unoebauer@mpa-garching.mpg.de}
and S.~A.~Sim$^2$\\
  $^1$Max-Planck-Institut f\"ur Astrophysik, Karl-Schwarzschild-Str.~1, 85748
  Garching, Germany\\
  $^{2}$Astrophysics Research Centre, School of Mathematics and Physics, Queen's
  University Belfast, Belfast BT7 1NN, UK\\
}

\begin{document}
\maketitle
\begin{abstract}
  Radiative pressure exerted by line interactions is a prominent driver of
  outflows in astrophysical systems, being at work in the outflows emerging from
  hot stars or from the accretion discs of cataclysmic variables, massive young stars
  and active galactic nuclei. In this work, a new radiation hydrodynamical approach to model
  line-driven hot-star winds is presented. By coupling a Monte Carlo radiative
  transfer scheme with a finite-volume fluid dynamical method, line-driven mass
  outflows may be modelled self-consistently, benefiting from the advantages of
  Monte Carlo techniques in treating multi-line effects, such as multiple
  scatterings, and in dealing with arbitrary multidimensional configurations. In
  this work, we introduce our approach in detail by highlighting the key
  numerical techniques and verifying their operation in a number of simplified
  applications, specifically in a series of self-consistent, one-dimensional,
  Sobolev-type, hot-star wind calculations. The utility and accuracy of our
  approach is demonstrated by comparing the obtained results with the
  predictions of various formulations of the so-called \cak theory and by
  confronting the calculations with modern sophisticated techniques of
  predicting the wind structure. Using these
  calculations, we also point out some useful diagnostic capabilities our approach
  provides. Finally we discuss some of the current limitations of our method, some
  possible extensions and potential future applications.
\end{abstract}
\begin{keywords}
  methods: numerical -- hydrodynamics -- radiative transfer -- stars: winds
\end{keywords}

\section{Introduction}

Predicting massive star evolution is dramatically complicated by the presence of
powerful winds, constituting an important mass-loss mechanism \citep[see, for
example, overview in][]{Kudritzki2000}. The evolutionary timescales of the star
and its luminosity are, for example, significantly affected by this continuous
mass loss. In addition to influencing the stars themselves, these winds also
affect their environments by injecting energy and momentum into the interstellar
medium and supplying chemically enriched material. For a detailed study of all
these processes and effects a firm understanding of the mechanisms driving
stellar winds is required.

In the case of hot O and B stars, the main driver for the mass outflow has been
identified as the momentum transfer mediated by a multitude of atomic line
interactions of the radiation field in the wind material \citep{Lucy1970,
Castor1975}.  Over the years, this picture has been significantly refined and a
``standard model'' \citep[see, for example, review by][]{Puls2008} for
line-driven winds has emerged, with important contributions from
\citet{Castor1975,Pauldrach1986,Kudritzki1989}. With the inclusion of multi-line
effects, in particular by \citet{Vink2000}, this model has become very
successful in describing hot-star winds and predicting their properties.
However, observational diagnostics have identified a number of puzzles
that challenge this ``standard model'' \citep[see again][]{Puls2008}. Among
these, the so-called \textit{clumping problem} takes a prominent role. Contrary
to the assumptions of the ``standard model'', hot-star winds seem to exhibit
strong temporal and spatial variability.  Intense effort has been invested into
understanding the properties and the nature of these clumpy outflows, both from
an observational and theoretical viewpoint. In light of these efforts, we have
developed a new radiation hydrodynamical approach to numerically model
line-driven mass outflows self-consistently. Due to its reliance on Monte Carlo
radiative transfer techniques, this approach should be well-suited for
time-dependent and multidimensional self-consistent studies of hot-star winds.
In previous works, a Monte Carlo radiative transfer scheme has already been
successfully coupled with a fluid dynamical method, resulting in the radiation
hydrodynamical approach \mcrh \citep{Noebauer2012, Noebauer2014}. In this study
we present extensions to \mcrh and demonstrate the utility of this method for
the investigation of line-driven winds. This work has been partially carried out
during the PhD project of \citet{Noebauer2014}.

The presentation of our approach is structured as follows: in Section
\ref{sec:theory}, we briefly recapitulate some basic concepts of hot-star winds
and outline different solution strategies to address this problem. We also
highlight the advantages of Monte Carlo radiative transfer techniques for the
study of line-driven mass outflows. We continue with a detailed description of
the numerical techniques relevant for our approach in Section
\ref{sec:numerics}. In Section \ref{sec:wind_sims}, we present a series of
simple wind simulations to test our approach and in Section \ref{sec:mv08_sims},
we confront our method with an alternative, modern and sophisticated technique
of predicting the wind structure. The obtained results and potential future
improvements of our method are discussed in detail in Section
\ref{sec:discussion}.

\section{Theory}
\label{sec:theory}

\subsection{Solving the stationary hot-star wind problem}
\label{sec:stat_wind}

Numerous techniques and approaches have been used to solve for the structure of
hot-star line-driven winds and derive their principle properties. Many of these
share the common feature of addressing the problem in a
``radiation-hydrostatics'' fashion. By assuming that the mass outflow is smooth,
spherically-symmetric and in a stationary state, the wind structure may be
derived from global energy conservation considerations (e.g.\
\citealt{Abbott1985}, \citealt{Lucy1993}, \citealt{Sim2004}) or by solving the
momentum equation alone (e.g.\ \citealt{Castor1975}, \citealt{Pauldrach1986},
\citealt{Kudritzki1989}). If included, inhomogeneities in the mass outflow,
i.e.\ clumps, are typically incorporated in a parameterised fashion, for
example, by introducing a clumping factor \citep[e.g.][]{Schmutz1997,Hamann1998}. In the simplified, stationary situation,
the momentum equation reduces to 
\begin{equation}
  u \frac{\mathrm{d} u}{\mathrm{d} r} + \frac{1}{\rho}
  \frac{\mathrm{d} P}{\mathrm{d}r} + \frac{G M_{\star}}{r^2} = g_{\mathrm{rad}},
  \label{eq:force_balance_wind}
\end{equation}
after introducing the wind velocity $u$, its density $\rho$, the thermodynamic
pressure $P$, the stellar mass $M_{\star}$ and the radiative acceleration
$g_{\mathrm{rad}}$. Solving this equation is still very challenging, since the bulk of the radiative
acceleration derives from the momentum transfer mediated by interactions of the
radiation field emitted by the central star with a large number of atomic line
transitions. The contribution due to Thomson scattering may be taken into account by
reducing the stellar mass in the gravity term by a factor $(1 -
\Gamma_{\mathrm{e}})$, using the Eddington factor with respect to electron
scattering, $\Gamma_{\mathrm{e}}$. 

Various approaches have been used to calculate the line-driving force. We
briefly sketch the basic principle of the so-called Castor--Abbott--Klein (\cak) theory
(after \citealt[][]{Castor1975}; see in particular \citealt{Abbott1982},
\citealt{Pauldrach1986}, \citealt{Kudritzki1989} for improvements of the
original approach) since it will be used for comparison in this work.
\citet{Castor1975} found an approximate analytic
expression for the line-driving force. In particular, the line-driving force is
expressed as a multiple of the electron-scattering contribution and this
so-called force-multiplier is approximated by a power-law parameterisation.
Given this form of the radiative acceleration, an analytic solution to the
momentum equation may be identified. By performing a critical point analysis, the
wind velocity is found to follow a $\beta$-type law
\begin{equation}
  u(r) = u_{\infty} \left(1 - \frac{R_{\star}}{r}\right)^{\beta}
  \label{eq:cak_velocity}
\end{equation}
with $\beta = 0.5$ in the original \cak case of \citet{Castor1975}, in which the
central star is assumed to radiate like a point source and with $\beta \approx
0.8$ if the finite extent of the star is taken into account
\citep[see][]{Pauldrach1986, Kudritzki1989}. Throughout this work, we reference
this improved formulation of the \cak approach as \mcak (``Modified-\cak''). The
mass-loss rate of the wind scales as 
\begin{equation}
  \dot M \propto C(\alpha) (k L_{\star})^{\frac{1}{\alpha}} \left[ M_{\star} (1 - \Gamma_{\mathrm{e}})
  \right]^{1 - \frac{1}{\alpha}}.
  \label{eq:mass_loss_rate_cak}
\end{equation}
In addition to the stellar parameters (luminosity $L_{\star}$, photospheric
radius $R_{\star}$ and central mass $M_{\star}$), the parameters of the
power-law fit, $k$ and $\alpha$, appear. The functional form of the coefficient
$C(\alpha)$ differs slightly between the \cak and \mcak approaches. More details
about $C(\alpha)$ and the \cak/\mcak methods, as used in this work, are provided in
Appendix \ref{app:cak}. 

A significant advantage of these radiation-hydrostatic techniques lies in the
analytic expressions for the wind structure they either provide or in the
possibility to efficiently solve the problem numerically \citep[see,
e.g.,][]{Abbott1985, Pauldrach1994, Vink1999, Puls2005}, even if more
sophisticated treatments of the line-driving force are included.  In both cases,
wind models may be constructed for entire grids of different stellar parameters
\citep[e.g.][]{Abbott1982,Vink2000, Puls2005, Muijres2012}. A drawback
lies in the omission of a self-consistent treatment of temporal and spatial
variations in the wind flow.

\subsection{A dynamical approach}

To account for, follow and solve the temporal and spatial variability of stellar
winds, the full radiation hydrodynamical problem the radiatively-driven mass
outflow constitutes has to be addressed
\citep[e.g.][]{Owocki1988,Feldmeier1995,Feldmeier1997,Dessart2003,Dessart2004,Dessart2005}.
In such an approach, a numerical solution to the full radiation hydrodynamical
equations describing the conservation of mass, momentum and energy has to be
formulated. In contrast to a pure hydrodynamical problem, the energy and
momentum transfer between the radiation field and the fluid material are
included. These terms are found by solving the co-evolution of the radiation
field in parallel. However, due to the complexity of the full radiation
hydrodynamical problem, a simplified treatment of the radiation field and
radiative transfer is typically adopted
\citep[c.f.][]{Owocki1988,Feldmeier1995,Dessart2003}. The radiation hydrodynamical equations
are given in Appendix \ref{app:rh_equations} in the form in which they are used
in this study.

In this work we propose a new numerical approach to perform such dynamical
self-consistent simulations of hot-star winds. The distinguishing feature of our
method is the incorporation of Monte Carlo radiative transfer techniques. In
particular, we build upon the general-purpose Monte Carlo radiation
hydrodynamics scheme \mcrh, developed by \citet{Noebauer2012,Noebauer2014} and
adapt it to the line-driven wind environment. Other Monte Carlo-based radiation
hydrodynamical techniques have been developed by \citet{Nayakshin2009},
\citet{Acreman2010}, \citet{Haworth2012} and most recently by \citet{Roth2015},
who place a particular emphasis on the incorporation of implicit Monte Carlo
methods \citep[see][]{Fleck1971}, and by \citet{Harries2015}.

\subsection{Advantages of Monte Carlo techniques}

Monte Carlo techniques have become established as a competitive and successful
numerical approach to solve radiative transfer problems due to the increasing
availability of computational resources and the more and more levels of
sophistication added to the formalism (see specifically, \citealt{Abbott1985},
\citealt{Lucy1999}, \citealt{Lucy2002}, \citealt{Lucy2003},
\citealt{Carciofi2006}). The appeal of using Monte
Carlo techniques derives from their advantages in treating interaction
physics (see, for example, \citealt{Pozdnyakov1983} for Monte Carlo calculations
of comptonisation or \citealt{Kasen2006} and \citealt{Kromer2009}, for
presentations of fully-fledged three-dimensional Monte Carlo codes to solve
multi-frequency radiative transfer in Type Ia supernova ejecta) and from the
ease with which arbitrary geometrical configurations are treated (see, for
example, \citealt{Camps2013}, describing a Monte Carlo radiative transfer scheme
on a Voronoi mesh). These advantages all derive from the microscopic viewpoint
that Monte Carlo techniques take by solving the propagation history of a large
number of representative photons probabilistically. They are very
relevant for the study of hot-star winds, in which a multitude of atomic line
interactions are the main driver, multiple scatterings frequently occur and a
clumpy irregular structure of the mass outflow is expected. Thus, Monte Carlo
techniques have been frequently used to address the radiative transfer problem
in hot-star winds, for example by \citet{Abbott1985}, \citet{Lucy1993},
\citet{Schmutz1997} \citet{Vink1999}, \citet{Vink2000}, \citet{Sim2004},
\citet{Mueller2008}, \citet{Muijres2012}, \citet{Surlan2012}.

In this work, we aim at exploiting the advantages of Monte Carlo techniques for
solving the line-driving problem in a dynamical and self-consistent approach by
using a Monte Carlo radiation hydrodynamical technique. This strategy
distinguishes itself form previous Monte Carlo-based investigations by its
treatment of the coupled co-evolution of the radiation--wind state. Moreover,
the radiative acceleration is determined locally, which sets us apart from some
earlier Monte Carlo-based studies, such as \citet{Abbott1985}, \citet{Lucy1993}, \citet{Vink1999}
and \citet{Sim2004}, in which the momentum transfer onto the wind material was
derived from global considerations without insisting on local consistency
(however, see \citealt{Mueller2008}, for developments towards a
  consistent local force balance in Monte Carlo-based approaches).

\section{Numerical Methods}
\label{sec:numerics}

In order to adequately address the radiation--matter coupling problem in the
stellar wind environment, we have extended our numerical framework whose
underlying methodology and first implementation has been described in
\cite{Noebauer2012}. In the following, the basic principles
of our approach are briefly recalled and the new extensions described in
detail. 

\subsection{Basic operator-split approach}
\label{sec:basicapproach}

In our approach, a simple operator-splitting scheme is employed, coupling a
time-dependent Monte Carlo radiative transfer simulation with a finite-volume
fluid dynamical calculation. In the latter, the piecewise parabolic method (PPM,
\citealt{Colella1984}) is used to construct a series of Riemann problems at the
interfaces of the computational domain, which are solved with a standard Riemann
solver \citep[HLLC, see][]{Toro1994, Batten1997} to determine the mass, energy
and momentum flux through the interfaces and eventually the evolution of these
conserved quantities in the cells. In the following radiative transfer step, a
Monte Carlo simulation is performed to determine the evolution of the radiation
field and reconstruct the energy and momentum transfer between the radiation
field and the fluid material. In the final step of the splitting approach, the
momentum and energy of the fluid are altered in accordance with these transfer
terms.

This basic outline of the numerical scheme remains valid for the stellar wind
applications. Alterations and extensions of the scheme only concern the
inclusion of additional physical effects or adopted simplifications that are
relevant for an adequate treatment of the stellar wind environment. In
particular, the gravitational field originating from the central star is taken
into account, an isothermal treatment of fluid dynamics is adopted and the
radiative transfer scheme is generalised to incorporate frequency-dependent
resonant line interactions. These alterations are systematically introduced and
presented in detail in the following sections.

\subsection{Isothermal hydrodynamics and gravity}
\label{sec:isohydro}

In this work, we reduce the complexity of the hydrodynamical
calculations in our modelling approach of hot-star winds by adopting an
isothermal approximation. This simplification, which has been used in numerous
investigations of line-driven hot-star winds
\citep[e.g.][]{Abbott1982,Abbott1985,Owocki1988,Vink1999,Vink2000,Sim2004}, is
justified by the continuous irradiation of the wind material by the central
star, which together with a characteristic cooling time scale that is faster
than the flow time scale sustains the wind material roughly at the effective
temperature of the star \citep{Klein1978a}. In the isothermal approximation, only
the mass and momentum conservation equations have to be addressed in the fluid
dynamical calculation. With the equation of state of
an isothermal flow directly connecting the fluid density and pressure
\begin{equation}
  P = a_{\mathrm{iso}}^{2} \rho,
  \label{eq:isothermal_eos}
\end{equation}
a solution of the energy conservation equation is not required. In the case of
an ideal gas with constant temperature $T_{\mathrm{iso}}$ and mean molecular
weight $\mu$, the isothermal sound speed $a_{\mathrm{iso}}$ is
\begin{equation}
  a_{\mathrm{iso}} = \sqrt{\frac{k_{\mathrm{B}} T_{\mathrm{iso}}}{\mu}}.
  \label{eq:isothermal_sound_speed}
\end{equation}

We modify the fluid dynamical solution procedure in the isothermal version of
\mcrh by reconstructing only the primitive variables density and the velocity.
Using the isothermal equation of state, the pressure may be directly determined
and the Riemann problems at the cell interfaces solved. Here, we use a
particular version of the HLLC Riemann solver, adopted from the \athena
code\footnote{The source code of this finite-volume magneto-hydrodynamical code
  may be obtained from \url{https://trac.princeton.edu/Athena/}.}
  \citep{Stone2008}, which is tailored to the particular form of the simple
  waves in isothermal flows.

An important contribution to the momentum balance in hot-star winds derives from
the gravitational pull exerted by the central object
\begin{equation}
  g = -\frac{G M_{\star}}{r^2}.
  \label{eq:gravitional_field}
\end{equation}
Following \citet{Colella1984}, we incorporate the effect of the star's
gravitational pull into the isothermal fluid dynamical calculation by accounting
for the additional momentum density contribution
\begin{equation}
  \Delta p^n = \frac{1}{2} \Delta t (\rho^n g^n + \rho^{n+1} g^{n+1}),
  \label{eq:gravity_effect_on_p}
\end{equation}
in the final determination of the new fluid dynamical state at the end of the
splitting step. Here, the superscripts $n$ and $n+1$ mark quantities at the
beginning and the end of a time step with length $\Delta t = t^{n+1} - t^n$.

\subsection{Monte Carlo principles}
\label{sec:mcprinciples}

At the heart of the Monte Carlo radiative transfer machinery lies the
discretization of the radiation field into a large number of quanta, so called
\textit{packets}, which are launched into the computational
domain. In a series of random number experiments, the packets describe
radiation--matter interactions and the temporal evolution
of the radiation field. The basic layout of our Monte Carlo method has already
been presented in detail in \cite{Noebauer2012}. Here, we only recall some
important aspects and dedicate the bulk of the presentation to the additions
related to the frequency-dependent calculation.

\subsection{Discretization, initialization and propagation}
\label{sec:mcdiscretization}

We adopt the indivisible energy packet scheme of \citet{Abbott1985} and
\citet{Lucy1999}. Thus, Monte Carlo packets are interpreted primarily as parcels
of radiative energy. To model the continuous illumination of the outflowing
material by the central star, a number of packets are injected into the
computational domain through the lower boundary in each radiative transfer step,
Assuming that the star radiates as a black body with effective temperature
$T_{\mathrm{eff}}$,
\begin{equation}
  N_{\mathrm{inject}} = \frac{4 \pi R_{\star}^2 \sigma_{\mathrm{R}} T_{\mathrm{eff}}^4 \Delta
  t}{\varepsilon}
  \label{eq:number_of_inflow_packets}
\end{equation}
new packets, each with energy $\varepsilon$, are launched from the inner
boundary (located at $R_{\star}$)
during a time step of length $\Delta t$.  Neglecting any limb-darkening effect,
the initial propagation direction of these packets follows 
\begin{equation}
  \mu = \sqrt{z}.
  \label{eq:mu_no_limb_darkening}
\end{equation}
Here and in the following, $\mu$ denotes the cosine of the directional angle
between the trajectory and the symmetry axis and $z$ represents a random number, uniformly drawn from
the interval $[0, 1]$.
The frequency assignment of the packets is chosen to reflect the Planck
function\footnote{See for example \citet{Carter1975},
for an efficient algorithm to sample this function.}
\begin{equation}
  B(\nu, T) = \frac{2 h \nu^3 }{c^2} \frac{1}{e^{h \nu / k_{\mathrm{B}} T} - 1}.
  \label{eq:planckfunction}
\end{equation}
This frequency assignment is adopted for this current study for simplicity and
convenience. Sampling a realistic atmosphere model, however, does not pose a
conceptual or implementation challenge \citep[see, e.g.,][]{Abbott1985}.

After determining the initial properties of the packets, they are launched and initiate the
propagation process.
As detailed in \citet{Noebauer2012}, we follow a mixed-frame approach and
perform the packet propagation procedure in the lab frame (\textit{LF}) in which
the computational grid is defined. However, all interaction processes are
treated in the local co-moving frame (\textit{CMF}), in which the fluid is at
rest and where the material functions take convenient forms. Transformation
rules relate quantities between these two frames. These are derived in detail,
for example, in \citet{Mihalas1984} and are listed partially in
\citet{Noebauer2012}.

During their propagation, packets interact whenever they have covered
a pathlength equivalent to the optical depth 
\begin{equation}
  \tau =  -\ln z,
  \label{eq:opticaldepthsampling}
\end{equation}
which is determined for every packet in a random number experiment.  The
procedure with which the nature of the interaction event is determined is
presented in the next section. Once the details of the interaction process are
established, the packet properties are updated accordingly. We highlight the
case of line interactions, which we treat as resonant scatterings in the
\textit{Sobolev approximation} (after \citealt{Sobolev1960}; see also, for
example, \citealt{Lamers1999} for a summary of this approximation and
\citealt{Pauldrach1986} for a discussion of its applicability to hot-star
winds). In this case, the packet is assigned a new propagation direction in the
CMF according to the Sobolev escape probability
\begin{equation}
  p(\mu) \propto \frac{1 - \exp(-\tau_{\mathrm{s}})}{\tau_{\mathrm{s}}}.
  \label{eq:sobolev_escape_probability}
\end{equation}
Here, $\tau_s$ denotes the Sobolev optical depth, which will be explicitly
introduced in Section \ref{sec:ionisation} and which depends on the direction of
the photon trajectory. The changes in the LF frequency and energy of the packet
follow from the Doppler effect and from energy conservation in the CMF. Assuming
that the process of sampling the Sobolev escape probability returned the CMF
propagation direction $\mu^{\mathrm{a}}_0$  and accounting for the appropriate
transformation laws, the resonant line scattering process results in the new
packet quantities:
\begin{align}
  \mu^{\mathrm{a}} &= \frac{\mu_0^{\mathrm{a}} + \beta}{1 + \beta \mu_0^{\mathrm{a}}},
  \label{eq:scattering_direction}\\
  \varepsilon^{\mathrm{a}} &= \gamma^2 (1 - \beta \mu^{\mathrm{b}}) (1 + \beta
  \mu_0^{\mathrm{a}}) \varepsilon^{\mathrm{b}},
  \label{eq:scattering_energy}\\
  \nu^{\mathrm{a}} &= \gamma^2 (1 - \beta \mu^{\mathrm{b}}) (1 + \beta
  \mu_0^{\mathrm{a}}) \nu^{\mathrm{b}}.
  \label{eq:scattering_frequency}
\end{align}
Here and in the following, a subscribed ``0'' denotes CMF quantities and the
superscripts $^{\mathrm{b}}$ and $^{\mathrm{a}}$ distinguish packet properties \textit{before} and
\textit{after} the interaction process. Moreover, the standard notation of
special relativity is adopted, introducing $\beta = u/c$ and $\gamma = 1 /
\sqrt{1 - \beta^2}$. Above and in the following we keep full account of the
relativistic terms and only in the final expression reduce the accuracy to
$\mathcal{O}(u/c)$, which is appropriate for modelling hot-star winds. 

In addition to the physical radiation--matter interactions, packets also
experience so-called \textit{numerical events} due to the spatial and temporal
discretization of the computational domain. When a packet intercepts a grid cell
boundary, some packet properties, such as the optical depth distance to the next
interaction, have to be re-determined to be compatible with the physical
conditions in the target cell. At the end of the time step, the trajectories of
all remaining packets are interrupted and their properties stored to
allow them to resume their propagation during the next simulation cycle.

\subsection{Optical depth summation}
\label{sec:optical_depth_summation}

The sole purpose of the packet propagation process lies in the determination of
packet interaction histories from which the radiation field characteristics can
be reconstructed. For the stellar wind application, this concerns primarily the
line-driving force. Crucial for the determination of the packet trajectories is
the decision about where and how the packets interact. As already mentioned,
packets experience physical radiation--matter interactions after having
accumulated an integrated optical depth equal to the random number experiment
outcome (\ref{eq:opticaldepthsampling}). For the current work, the wealth of
possible interaction processes has been restricted to include only
frequency-dependent bound-bound processes, which we treat as resonant scatterings
in the Sobolev limit (see previous section). All continuum processes are
neglected, apart from Thomson scattering, which is incorporated approximately by
reducing the mass of the central star as outlined in Section
\ref{sec:stat_wind}.

Once a packet interacts, the location of this event in optical depth space has
to be translated into a physical position within the computational domain. This
procedure is trivial in the grey case, since optical depth and pathlength only
differ by the opacity, a constant multiplicative factor in the CMF, but
complicated in the presence of frequency-dependent processes. We adopt a
simplified version of the technique of \citet{Mazzali1993} to locate the
line-interaction events packets perform. On its trajectory, a packet propagates
freely to the Sobolev point of the next line with which it comes into resonance.
Each time such a resonance point is reached, the optical depth is incremented by
the full line optical depth of the corresponding transition. The packet
undergoes an interaction once the value drawn in (\ref{eq:opticaldepthsampling})
is surpassed by the optical depth accumulated.  If this occurs during the
instantaneous increases at one of the Sobolev points, the packet undergoes a
resonant line interaction, otherwise it may leave the current grid cell
uninterrupted. This procedure may be easily extended to include additional
interaction types, in particular frequency-independent processes, such as
Thomson scatterings \citep[see][]{Mazzali1993}, but for the current work we have
omitted to do so.

\subsection{Monte Carlo estimators}
\label{sec:estimators}

In reconstructing the radiation field characteristics from the ensemble of
packet interaction histories, we follow the volume-averaged estimator approach
proposed by \cite{Lucy1999} and refined by \cite{Lucy2003, Lucy2005}.  This
formalism aims at reducing the statistical fluctuations inherent to the Monte
Carlo approach by increasing the number of contributions to the packet census.
For the case of frequency-independent processes being the only interaction
channel, adequate estimators have already been derived and presented by
\cite{Noebauer2012} and similarly by \citet{Roth2015}.

To determine the radiative acceleration due to spectral line interactions, we
consider the momentum transferred in such an event. Assuming that these
interactions occur as resonant scatterings, a packet transfers 
\begin{equation}
  \Delta p = \frac{\varepsilon^{\mathrm{b}}}{c} \left[ \mu^{\mathrm{b}} -
    \gamma^2 (\mu_0^{\mathrm{a}} +
  \beta)(1 - \beta \mu^{\mathrm{b}}) \right]
  \label{eq:momtransfer}
\end{equation}
momentum onto the material. Estimators for the radiation force
can be obtained by summing over the transfer terms of all interacting
packets. To reduce the statistical fluctuations in these estimators we follow the
suggestion of \cite{Lucy1999a} and include all packets that come into resonance
with a line and weight their contributions with the corresponding interaction
probability given by $(1 - e^{-\tau_{\mathrm{s}}})$. Taking the forward-backward symmetry of the re-emission
into account, thus cancelling all terms that are of odd power in
$\mu_0^{\mathrm{a}}$, the
following estimator for the radiation force [c.f.\ Equation
(\ref{eq:radiation_force_1})] due to line interactions are
obtained:
\begin{equation}
  G^1_{\mathrm{line}} = \frac{1}{\Delta V c \Delta t} \sum (1 -
  e^{-\tau_{\mathrm{s}}})
  \varepsilon (\mu - \beta).
  \label{eq:g3lineestimator}
\end{equation}
Here, the volume of a grid cell, $\Delta V$ appears. The superscript
$^{\mathrm{b}}$ has
been dropped and only terms of $\mathcal{O}(u/c)$ have been retained. Using this
estimator, the radiation force is calculated and the fluid momentum updated in
the final splitting step.

\subsection{Ionization and level population}
\label{sec:ionisation}

The strength of the different line transitions, as encoded in the corresponding
values for the Sobolev optical depth, $\tau_{\mathrm{s}}$, depends on the excitation and
ionisation balance in the wind. Thus, the Monte Carlo radiative transfer routine
requires a separate calculation step which determines the ionisation and the
level population in each cell of the computational grid. In this first study, we
follow \citet{Abbott1985}, \citet{Vink1999} and \citet{Sim2004} and adopt an
approximate non-LTE treatment. We stress, however, that nothing in our radiative transfer
or hydrodynamics formalism requires these approximations. A full, non-LTE scheme
for calculating ionisation and excitation states could be incorporated following
the methods described by \citet[][see Section \ref{sec:nonlte}]{Lucy2003}.

Following our simplified strategy, we determine the ionisation balance by applying the
modified nebular approximation \citep[see, e.g.][]{Mazzali1993},
\begin{align}
  \frac{N_{j+1}}{N_j} &= \left( \frac{N_{\mathrm{e}}}{W} \right)^{-1}\nonumber\\
  &\times [ \zeta_j + W (1 - \zeta_j)]
  \sqrt{\frac{T_{\mathrm{e}}}{T_{\mathrm{R}}}} \left( \frac{N_{j+1} N_{\mathrm{e}}}{N_j}
  \right)^{\ast}_{T_{\mathrm{R}}}.
  \label{eq:modifiednebular}
\end{align}
This expression relates the number densities of two successive ion stages,
$N_j$, $N_{j+1}$ with the electron number density $N_{\mathrm{e}}$. Compared to
a pure LTE calculation based on the Saha equation, whose results are denoted by
the asterisks, modifications due to the dilution of the radiation field and due
to recombination effects are taken into account. As a consequence, the dilution
factor $W$, the ratio of the electron and radiation temperatures,
$T_{\mathrm{e}}$ and $T_{\mathrm{R}}$, and the fraction $\zeta_j$ of
recombination processes going directly to the ground state appear in the above
formulation.

In the determination of the level population, we again follow \cite{Abbott1985}
and identify levels with no permitted electromagnetic dipole transitions to
lower energy levels as metastable; these levels are assumed to obey Boltzmann
statistics
\begin{equation}
  \left( \frac{n_i}{n_1} \right) = \left( \frac{n_i}{n_1}
  \right)^{\ast}_{T_{\mathrm{R}}},
  \label{eq:metastablelevels}
\end{equation}
with ``1'' denoting the ground state.  For all other levels, the effect of
radiative processes on the level population is crudely taken into account by
including a dilution factor \citep[see][]{Vink1999, Sim2004}
\begin{equation}
  \left( \frac{n_i}{n_1} \right) = W \left( \frac{n_i}{n_1}
  \right)^{\ast}_{T_{\mathrm{R}}}.
  \label{eq:nonmetastablelevels}
\end{equation}

For the current work, we do not solve Equation (\ref{eq:modifiednebular}) iteratively to
determine the ionisation state. Instead, we use a predefined characteristic
electron density $N_{\mathrm{e}}/W$ and calculate the ionisation accordingly \citep[see,
e.g.][for a comparable strategy]{Abbott1982} in all our wind simulations. Also,
the radiation temperature is treated in a simple fashion according to the
prescription 
\begin{equation} T_{\mathrm{R}} = T_{\mathrm{e}} =
  T_{\mathrm{eff}} 
  \label{eq:stellar_winds_temperature}
\end{equation}
rather than relying on predictions of realistic atmosphere models. 

With the excitation and ionisation balance accessible through Equations (\ref{eq:modifiednebular}), (\ref{eq:metastablelevels}) and
(\ref{eq:nonmetastablelevels}) in each cell of the
computational grid, the Sobolev optical depth of a line transition encountered
on a packet trajectory along the direction $\mu$ may be calculated according to 
\begin{align}
  \tau_{\mathrm{s}} &= \frac{c}{\nu_0} \frac{(\kappa_l \rho)_{r_{\mathrm{s}}}}{\left[(1 - \mu^2) \frac{u}{r}
  + \mu^2 \frac{\mathrm{d}u}{\mathrm{d r}}\right]_{r_{\mathrm{s}}}}
  \label{eq:sobolev_optical_depth},\\
  \kappa_l \rho &= \frac{\pi e^2}{m_{\mathrm{e}} c} f_l n_l \left( 1 - \frac{n_u }{n_l}
  \frac{g_l}{g_u} \right).
  \label{eq:sobolev_optical_depth_kappa}
\end{align}
Here, $\nu_0$ denotes the rest frame frequency of the transition and $f_l$ its
oscillator strength. Also, the statistical weights $g_{l,u}$ associated with the
lower and upper energy levels connected by the transition and the electron
charge and mass, $e$ and $m_{\mathrm{e}}$, appear. Finally, we stress again that the
above expressions only depend on the physical conditions at the Sobolev point,
an essential feature which is highlighted by the subscribed $r_{\mathrm{s}}$.

\subsection{Velocity interpolation}

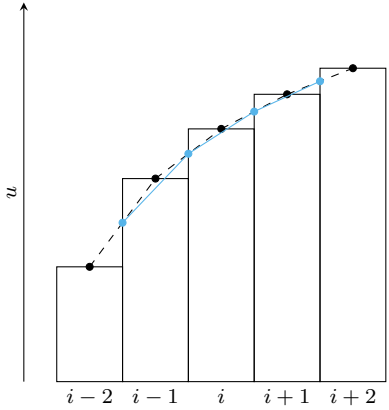
\begin{figure}
  \centering
  \input{./velocity_interp_cell.tikz}
  \caption{Illustration of the linear interpolation scheme to ensure continuity
  in the CMF packet frequency. The grid cell boundaries are treated as locations
where the fluid velocity and thus the CMF frequency is known at all
times. The fluid flow speed there is obtained by linear interpolation (dashed
line) between the volume-averaged values of the finite-volume scheme. The
velocity in the grid cells is then obtained by another linear interpolation step
between the boundary values (solid blue line).}
  \label{fig:vel_interp_cell}
\end{figure}

As photons propagate through the wind material, their CMF frequency is
continuously Doppler-shifted by the varying wind velocity. To reproduce this
behaviour in the Monte Carlo simulation, an interpolation scheme is required to
reconstruct the evolution of the wind velocity within the computational grid
cells. For the current work, we do not expect the formation of shocks but the
presence of a smoothly varying velocity field. Thus, in constructing the
algorithm, an artificial introduction of discontinuous jumps in the interpolated
wind velocity should be avoided: otherwise, packets may skip frequency regions
which may be populated by line transitions.  We prevent this by devising an
interpolation scheme that insists on continuity in the reconstructed wind
velocity at the interfaces between grid cells. First, the velocity at the
interfaces are determined by linear interpolation between the cell-averaged
values provided by the finite-volume fluid dynamical calculation.  The interface
values are then used in a second linear interpolation step to determine the
evolution of the wind velocity within the cell. Figure \ref{fig:vel_interp_cell}
graphically illustrates this procedure. 

Similarly to the spatial considerations, care has be taken when packets reach
the end of the time step. In order to ensure consistency in the CMF frequency
between successive time steps, the velocity is also interpolated linearly in
time between the results of two preceding fluid dynamical splitting steps.
Otherwise the changes of the fluid velocity due to fluid dynamics and the
radiation--matter coupling could introduce discontinuous jumps in the CMF
frequency.

\subsection{Boundary conditions}
\label{sec:wind_boundary_conditions}

In the hot-star wind calculations, the inner boundary of the computational
domain is set to the stellar surface, $R_{\star}$. So-called \textit{ghost
cells}, which are required for the fluid dynamical reconstruction step at the
domain boundaries, are inserted below this radius. The fluid state in these
inner ghost cells is found by using a temporally constant hydrostatic density
stratification and by extrapolating the velocity field from the innermost region
of the domain into the ghost cells using a low-order polynomial. For most wind
calculations presented in this work, the hydrostatic profile has been normalised
(arbitrarily) to $\rho_0 = \SI{e-11}{g.cm^{-3}}$. However, no significant
changes in the calculated wind structure was found if this value was modestly
varied. In the future, predictions from atmospheric models may be used for this
process instead.  With the wind density fixed but the possibility of the
velocity in the ghost cells to float, the mass flow through the boundary may
quickly adjust itself to the conditions in the wind.  This situation closely
resembles the boundary conditions presented by \citet{Owocki1988} for their
time-dependent studies.

As detailed in Section \ref{sec:mcdiscretization}, the inner edge of the domain
constitutes an inflow boundary for the radiation field. Any Monte Carlo packet
that is back-scattered through the inner boundary is discarded from the ensuing
propagation process. 

The outer edge of the computational domain is transparent to the radiation
field. All escaping packets are recorded to construct spectra or other
diagnostics during the postprocessing steps. Regarding the fluid state, the
outer boundary simulates the outflow of wind material. To this end, the values
for the fluid variables in the ghost cells are found by extrapolating the wind
flow. To minimise the effect of stochastic fluctuations introduced by the Monte
Carlo radiative transfer step, a linear regression scheme is used.

This outline of the boundary treatment concludes the description of the
numerical techniques which are relevant for this study and which have been added
to \mcrh. The performance of all these extensions has been verified in
a series of test calculations. A detailed description of this procedure and the
different test problems used during this process may be found in
\citet{Noebauer2014}.

\section{Hot-star wind test calculations}
\label{sec:wind_sims}

We now aim at demonstrating the capability of our approach to self-consistently
solve the line-driving problem in hot-star winds by considering an idealised
representation of the problem. Specifically, we test the basic accuracy of our
method by exploring whether it produces results compatible with the \cak and
\mcak theory when comparable assumptions about the physical effects at work are
adopted (smooth flow, Sobolev approximation, etc.). At the same time, we
investigate how these simple calculations compare with simulations in which the
full details as outlined in Section \ref{sec:numerics} are incorporated.
Consequently, this series of wind simulations, constitutes the basic testing and
validation step before we confront our method with the technique of
\citet{Mueller2008} in Section
\ref{sec:mv08_sims}. Finally, in Section \ref{sec:discussion}, we discuss how some of
the simplifications and approximations adopted in the current work may be
relaxed and a more realistic description of the line-driving problem be achieved
within our numerical framework in the future.

\subsection{Atomic data}

In all wind calculations, we assume that the material has solar
composition. We include the elements listed in Table \ref{tab:elem_ions} and
adopt the corresponding abundances of \citet{Asplund2009}.
\begin{table}
  \centering
  \caption{Overview of all chemical elements, that
    are taken into account in our wind calculations. For each species, the
    included ionisation stages are listed as well.}
  \begin{tabular}{cccc}
    \hline
    Element & Ions & Element & Ions \\
    \hline
    H  & I,II & He & I-III \\
    C  & I-IV & N  & I-VI  \\
    O  & I-VI & F  & I-VI  \\
    Ne & I-VI & Na & I-VI  \\
    Mg & I-VI & Al & I-VI  \\
    Si & I-VI & P  & I-VI  \\
    S  & I-VI & Cl & I-V   \\
    Ar & I-V  & K  & I-V   \\
    Ca & I-VI & Ti & I-VI  \\
    Cr & I-VI & Mn & I-VI  \\
    Fe & I-VI & Co & I-VI  \\
    Ni & I-VI &    &       \\
    \hline
  \end{tabular}
  \label{tab:elem_ions}
\end{table} 
Table \ref{tab:elem_ions} also identifies the ionisation stages that are
taken into account for each element. Information about the atomic structure and the line
transitions in these ions are taken from the same database used in the
study of colliding winds by \citet{Parkin2013}, which is based on the
\citet{Kurucz1995} atomic data set. To reduce the computational effort in the
Monte Carlo radiative transfer steps, we do not account for all line transitions
recorded in the data set, but only those with $\log g f > -6$.
We have explicitly verified that the inclusion of more weak lines does not
affect the outcome of our simulations. When required, the recombination fractions for the
included ions, $\zeta_i$, are adopted from the \python code \citep{Long2002}.

\subsection{Stellar parameters}

We perform all wind calculations for fixed sets of stellar parameters. For the
simulations series constituting the basic testing process, we consider a system
that is similar to the well-studied O-star $\zeta$-Puppis. The basic stellar
parameters for these calculations are listed in Table \ref{tab:zetapup_pars} and
are adopted from \citet{Puls1996}. 
\begin{table}
  \centering
  \caption{Properties of $\zeta$-Puppis according to \citet{Puls1996}. We adopt
  these stellar parameters in our wind test calculations.}
  \begin{tabular}{cc}
    \hline
    Parameter & Value\\\hline
    $M_{\star}$ & $52.5\, \mathrm{M}_{\odot}$\\
    $L_{\star}$ & $10^6 \mathrm{L}_{\odot}$\\
    $T_{\mathrm{eff}}$ & $\SI[retain-unity-mantissa=false]{4.2e4}{K}$\\
\hline
  \end{tabular}
  \label{tab:zetapup_pars}
\end{table}
These values imply a stellar radius of $R_{\star} = \SI{1.317e12}{cm}$.
Additionally, we use $N_{\mathrm{e}} / W =
\SI{e15}{cm^{-3}}$ in the $\zeta$-Puppis
calculations.  Performing an ionisation and excitation calculation as outlined
in the previous section, we find a mean electron scattering cross section of
$\sigma_{\mathrm{e}} = \SI{0.34}{cm^2.g^{-1}}$ throughout the wind, which
corresponds to the Eddington factor $\Gamma_{\mathrm{e}} = 0.5$. In all \mcrh
wind simulations presented in the following, we account for the effect of
electron scattering by reducing the stellar mass by the factor $(1 -
\Gamma_{\mathrm{e}})$, as described in Sections \ref{sec:stat_wind} and
\ref{sec:optical_depth_summation}.

\subsection{\cak fitting}
\label{sec:fitting}

\begin{figure}
  \begin{center}
    \includegraphics[width=84mm]{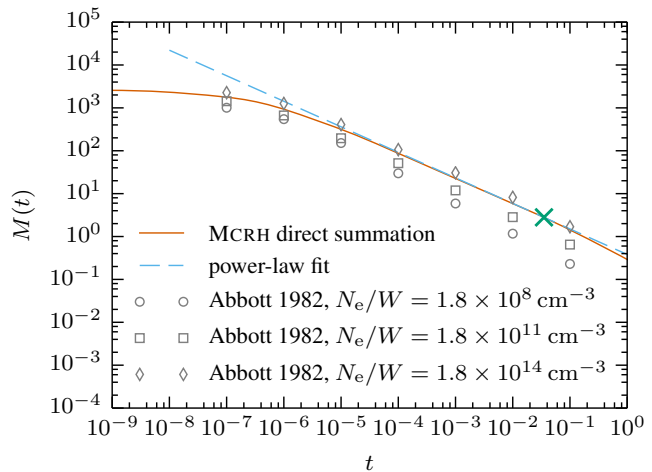}
  \end{center}
  \caption{Results of the direct determination of the
    force multiplier as a function of the
    optical depth parameter $t$, performed with \mcrh (orange) according to
    Equation (\ref{eq:force_multiplier_explicit_sum}). In the transition region, between the optically
    thick and thin regimes, a power-law fit according to Equation (\ref{eq:app_cak_mt})
    has been performed (dashed blue). The parameters of this fit are included
    in Table \ref{tab:cak_fit_results} and the predicted wind state (in terms of
    $t$), according to the \cak theory and the obtained $k$ and $\alpha$ values,
    is marked by a green cross. For comparison, the force multiplier values of
    \citet{Abbott1982}, determined for a star with $T_{\mathrm{eff}} =
    \SI{4e4}{K}$ and $N_{\mathrm{e}}/W = 1.8 \times 10^{8},\, 1.8 \times 10^{11}, \,
    \SI{1.8e14}{cm^{-3}}$ are included (grey open symbols).}
  \label{fig:cak_fit_results}
\end{figure}

In the \cak theory, the wind structure may be readily determined from the
stellar properties and the power-law parameters $k$ and $\alpha$ of the force
multiplier.  However, the wind characteristics, mass-loss rate and terminal wind
speed, are sensitive to the exact values of these parameters. Given our aim at
assessing the utility of our approach for solving the line-driving problem, we
refrain from consulting previous studies, such as \citet{Abbott1982} or
\citet{Pauldrach1987}, which provide these parameters for a wide range of
stellar conditions. Instead, we use the Monte Carlo routine of \mcrh to
determine the values for $k$ and $\alpha$. This way, we ensure that the
\mcrh--\cak/\mcak comparisons are not obscured by differences in atomic data
sets, the ionisation and excitation treatments or the stellar parameters. In
particular, we determine the values for $k$ and $\alpha$ by carrying out the
force-multiplier summation \citep[c.f.][]{Abbott1982}
\begin{equation}
  M(t) = \sum \frac{F_{\nu_0} \Delta \nu_{\mathrm{D}}}{F} \frac{1}{t} \left[
  1 - \exp\left(  {-\frac{\kappa_l t}{\Delta \nu_{\mathrm{D}}
  \sigma^{\mathrm{ref}}_{\mathrm{e}}}}\right)
\right]
  \label{eq:force_multiplier_explicit_sum}
\end{equation}
explicitly with the \mcrh modules for a large number of different values of the
dimensionless optical depth and by performing the power-law fit according to
Equation (\ref{eq:app_cak_mt}) afterwards. The Doppler width, $\Delta
\nu_{\mathrm{D}} = \nu_0 u_{\mathrm{th}}/c$, according to the thermal motion
$u_{\mathrm{th}}$ [see Equation (\ref{eq:app_cak_vth})], is used together with a
reference specific electron scattering cross section
$\sigma_{\mathrm{e}}^{\mathrm{ref}}$ [see Equation (\ref{eq:app_cak_line})]. The
results of the direct summation procedure are shown in Figure
\ref{fig:cak_fit_results} together with the fitting curve. 
As a reference, the results of \citet{Abbott1982}, obtained for comparable
physical conditions are included. In general, both calculations agree, but
noticeable differences are present, which are most likely a consequence of our
simplified treatment of ionisation and excitation, differences in the atomic
data sets and of the use of realistic stellar atmosphere models instead of a
simple black body by \citet{Abbott1982}. Given the purpose of our calculations,
these differences are irrelevant but highlight the need for consistency between
the \mcrh and \cak calculations. The final \cak parameters, which are used
throughout this work, are listed in Table \ref{tab:cak_fit_results}.
\begin{table}
  \centering
  \caption{Overview of the main \cak parameters determined in
    the fitting process and used throughout this work. In addition, the wind
    properties, as predicted by the \cak theory according to these parameters
    are listed in the bottom rows. The adopted values for $N_{\mathrm{e}}/W$,
    $\rho_0$ and $\Gamma_{\mathrm{e}}$ are included for completeness.}
  \begin{tabular}{cc}
    \hline
    \cak Parameter/Wind Property & Value \\
    \hline 
    $k$ & $0.381$\\
    $\alpha $ & $0.595$\\
    $t$ & $0.035$ \\
    $M(t)$ & $2.80$ \\
    $\dot M$ & $4.51 \times 10^{-5}\,M_{\odot}\,yr^{-1}$\\
    $u_{\infty}$ & $\SI{881}{km.s^{-1}}$\\
    $N_{\mathrm{e}}/W$ & $\SI{e15}{cm^{-3}}$\\
    $\rho_0$ & $\SI{e-11}{g.cm^{-3}}$\\
    $\Gamma_{\mathrm{e}} $ & $0.502$\\
    \hline
  \end{tabular}
  \label{tab:cak_fit_results}
\end{table}
Based on the force multiplier parameters, the \cak predictions for the mass-loss
rate and the terminal wind speed are calculated and also provided in the table.
Including the finite-cone effect according to the \mcak approach increases the
terminal wind speed by a factor of $\sim 2.68$ to about $u_{\infty} \approx
\SI{2360}{km.s^{-1}}$ and reduces the mass-loss rate to $\dot M \approx 2 \times
10^{-5} \, M_{\odot}\,\mathrm{yr}^{-1}$. With these modifications, the terminal
wind speed is compatible with previous investigations of the wind of
$\zeta$-Puppis \citep[e.g.][]{Puls1996}. However, the density in our wind is too
high. Specifically, the mass-loss rate is about a factor of $3.5$ larger than
established by \citet{Puls1996}. Again, our simplified treatment of the wind
ionisation and excitation conditions is most likely the cause for these
discrepancies. We stress once more, however, that as long as we use the
simplified description of the wind conditions consistently in the \cak/\mcak
calculations and the \mcrh simulations, the conclusions drawn from comparing the
corresponding results are unaffected.

\subsection{General parameters}
\label{sec:num_parameters}

In the \mcrh simulations, the evolution of the wind material is followed until a
steady-state structure emerges. All calculations are carried out on a
non-uniform spherical mesh with $512$ cells, which span the region between
$1\,R_{\star}$ and $10\, R_{\star}$. The cell size increases outwards from $1.76
\times 10^{-3}\, R_{\star}$ at the stellar surface to $1.73 \times 10^{-1}\,
R_{\star}$ at the outer edge of the domain. During each time step, the incident
radiation field is discretised by $5 \times 10^4$ packets, which sample the
black-body spectrum in the wavelength range between $\lambda_{\mathrm{min}} =
228\,${\AA} and $\lambda_{\mathrm{max}} = 22800\,${\AA}. The lower edge
corresponds to the ionisation edge of He\,II. Any radiation more energetic than
this threshold is assumed to be rapidly removed by bound-free absorptions by
helium and consequently not included in our consideration \citep[see][for a
similar strategy]{Sim2004}. For the initial wind state in the \mcrh
calculations, we adopt a structure that is similar to the \cak/\mcak solution
but scaled up or down. By experimenting with other initial wind configurations,
we have explicitly verified that the final stationary wind solution, which is
found in the calculations, is insensitive to the details of the initial state.
Furthermore, we have also ensured that enough Monte Carlo packets are used in
the simulations. Increasing the number does not change the overall shape of the
radiative acceleration but only reduces the strength of the stochastic
fluctuations (see Figure \ref{fig:wind_line_cmp}).

\subsection{Simulation series layout}
\label{sec:simulation_layout}

Using the general parameters outlined in the previous sections, we perform a
series of \mcrh simulations of a $\zeta$-Puppis-like wind. In this series we
successively increase the level of physical detail (see Section
\ref{sec:testing_results}). For the first stage, the point-source approximation and an
unattenuated radiation field will be used, similar to \citet{Castor1975}. In the
second set of calculations, the finite extent of the star will be taken into
account (similar to \citealt{Friend1986}, \citealt{Pauldrach1986} and
\citealt{Kudritzki1989}). During these two first steps of the series, the
ionisation and excitation calculation is further simplified by setting all
recombination fractions $\zeta_i = 1$ and by assuming a Boltzmann excitation
formula [i.e.\ $W = 1$ in Equation (\ref{eq:nonmetastablelevels})]. Finally, in the last stage of the series, the assumption of an
unattenuated radiation field will be dropped. In this calculation, we then use
the full ionisation and excitation description as outlined in Section
\ref{sec:ionisation}, including recombination fractions and the geometric
dilution factor in the excitation balance. This leads to a mild variation of the
degree of ionisation in the wind. When applicable, the results of the
\mcrh calculations will be confronted with the predictions of the \cak and \mcak
theory. These comparisons are performed on the basis of the stationary wind
structure found in all calculations. With this in mind, we further reduce the
computational complexity in all \mcrh simulations by following all Monte Carlo
packets until they escape the wind during each time step. This is equivalent to
assuming a light propagation speed much larger than the fluid flow and
constitutes a common strategy in Monte Carlo radiative transfer approaches, for
example in \python \citep{Long2002} or \tardis \citep{Kerzendorf2014}. Adopting
this technique absolves us of the need to invest computational power into the
initial build-up of the radiation field. As long as we are solely interested in
the final stationary wind state, this simplification is justified. For future
calculations, which aim at investigating the dynamical behaviour of the wind
structure, this modification of the Monte Carlo radiative transfer scheme will
be dropped.

\subsection{\mcrh \cak/\mcak module}
\label{sec:cakmodule}

To better judge the outcome of the \mcrh--\cak/\mcak comparison, we include an
additional set of numerical calculations performed with our approach.  Instead
of relying on the Monte Carlo radiative transfer procedure to determine the
line-driving force, we incorporate the analytic power-law expressions of the
\cak/\mcak approach. In each simulation cycle, after the fluid dynamical and
central gravity substeps, the instantaneous value of the dimensionless optical
depth parameter $t$ is determined in each cell and the force multiplier $M(t)$
calculated accordingly. In this procedure, the central star may be treated
either as a
point source [see Equation (\ref{eq:app_cak_mt})], or its finite extent may be
taken into account [see Equations (\ref{eq:app_mcak_mt}) and
(\ref{eq:app_mcak_mt_df})]. To determine the
required instantaneous velocity gradient in each cell, we rely on the algorithm
by \citet{Fornberg1988} to construct finite differences on arbitrarily spaced
grids. We will refer to all numerical radiation hydrodynamical calculations
performed with this module as \cakrh and \mcakrh respectively.

\subsection{Results}
\label{sec:testing_results}

\subsubsection{Calculation in the point-source limit}
\label{sec:point-source}

In the first set of \mcrh calculations, the central star is approximated by a
point source and the radiation field is assumed to be unattenuated. Within the
Monte Carlo approach, both simplifications are realised by injecting the packets
only on radial trajectories, i.e.\ replacing the random number experiment
(\ref{eq:mu_no_limb_darkening}) by $\mu = 1$, and by not changing the
propagation direction in interaction events, i.e.\ $\mu_0^{\mathrm{a}} =
\mu_0^{\mathrm{b}}$.

\begin{figure*}
  \begin{center}
    \includegraphics[]{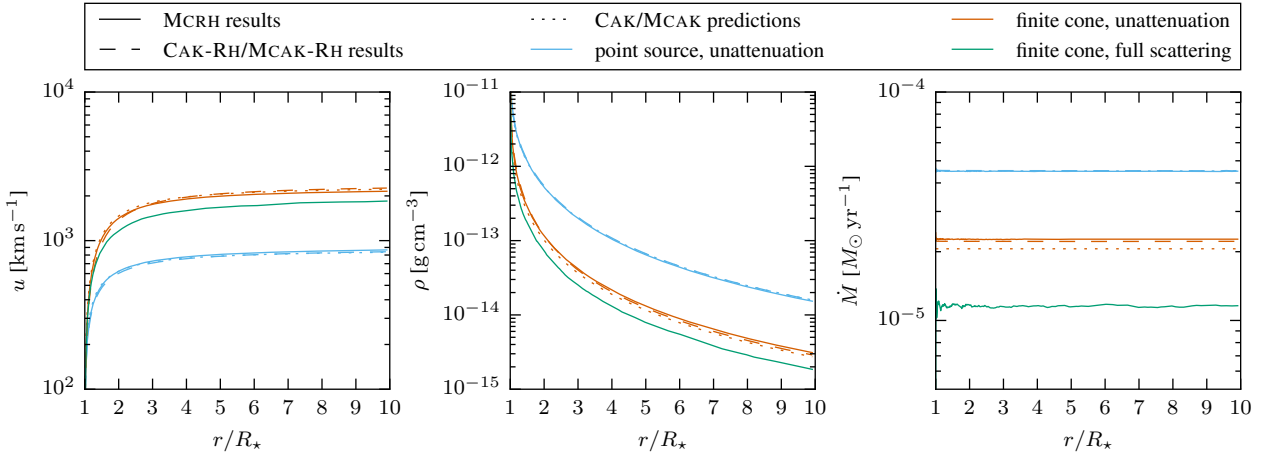}
  \end{center}
  \caption{Results of the stellar
    wind simulation series in terms of the final stationary wind velocity (left
    panel), density (central panel) and mass-loss rate (right panel). The
    colour-coding reflects the different stages of the simulation series. The
    calculations which are based on the point-source approximation and use the
    unattenuation of the radiation field are shown in blue. Red lines
    corresponds to calculations, in which the finite-cone effect is included and
    green to those which also include the full scattering procedure.
    All \mcrh results are presented by solid lines. Where applicable, the
    analytic predictions according to the \cak and the \mcak theory are included
  as dotted lines. In these cases, also the results obtained with 
  \cakrh/\mcakrh are given as an additional reference (dashed lines).}
  \label{fig:wind_cmp}
\end{figure*}

In Figure \ref{fig:wind_cmp}, the final stationary wind state, which establishes in the \mcrh
calculation, is shown in terms of the density, velocity and mass-loss rate.
The comparison with the results of the \cakrh simulation and the analytic
predictions according to the \cak theory shows a very good agreement. This
positive finding is a first indication for the accuracy and utility of our Monte
Carlo-based scheme to address the line-driving problem in hot-star winds
self-consistently.

\subsubsection{Including the finite-cone effect}
\label{sec:finite-cone}

During the second stage of the wind simulation series, the point-source
approximation is dropped and the finite extent of the central star is taken into
account, analogously to the \mcak approach of \citet{Pauldrach1986} and
\citet{Kudritzki1989}. In the \mcrh calculations during this stage, we allow
Monte Carlo packets to propagate on non-radial rays as well by sampling the
initial direction according to Equation (\ref{eq:mu_no_limb_darkening}).

As in the point-source calculations, the stationary wind structure obtained with
\mcrh is compared with the analytic predictions, now according to the \mcak
theory as outlined in Appendix \ref{app:cak}.  Again, we include the numerical
results calculated with the \mcakrh version of our scheme. When accounting for
non-radial photon propagation paths, the Monte Carlo-based results agree again very
well with the analytic predictions and the \mcakrh calculation, as illustrated
by Figure \ref{fig:wind_cmp}. 

\begin{figure}
  \begin{center}
    \includegraphics[]{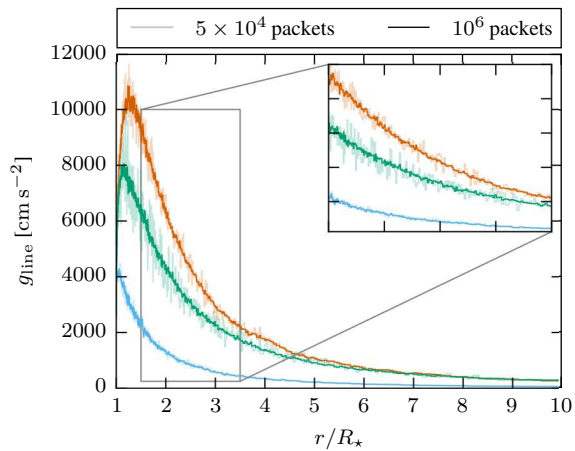}
  \end{center}
  \caption{The radial dependence of the line-driving force in the \mcrh
  calculations after a stationary state has established. The colour-coding is
  the same as in Figure \ref{fig:wind_cmp}. For the results shown by the weak solid 
lines, $5 \times 10^4$ packets where used to discretise the radiation field
emitted by the central star in each simulation cycle. For comparison, the full solid
lines show the corresponding acceleration, when the $10^6$ packets are used once
a stationary state has emerged. Notice the clear decrease in the Monte Carlo
noise but the same radial dependence of the radiative acceleration as in the
case with fewer packets.}
  \label{fig:wind_line_cmp}
\end{figure}

As expected from numerous previous studies, most notably from
\citet{Friend1986}, \citet{Pauldrach1986} and \citet{Kudritzki1989}, the
inclusion of the finite-cone effect leads to higher wind velocities, in our case
by a factor of $2.5$ in terms of the velocity at $r = 10\,R_{\star}$ compared to
the point-source case. At the same time, the mass-loss rate drops by a factor of
$0.5$. Figure \ref{fig:wind_line_cmp} illustrates the difference in the radial
dependence of the radiative acceleration, which is responsible for the
change in the structure of the wind flow.

\subsubsection{Full inclusion of scatterings}
\label{sec:res_att}

After having established the basic applicability of our approach to the
line-driving problem in the first stages of the simulation series, we take
another step towards a realistic description of the radiation field in hot-star
winds by dropping the unattenuation approximation and accounting for the full
scattering process in interactions. Now the entire procedure described in
Section \ref{sec:mcdiscretization} is performed: in particular, the emergent
propagation directions in atomic line interactions are drawn according to
Equation (\ref{eq:sobolev_escape_probability}). Accounting fully for the
multiple scattering phenomenon, these calculations reach beyond the capabilities
of the basic \cak and \mcak approaches (as outlined in Appendix \ref{app:cak}).
We emphasise, that the key consequence of multiple scattering lies in the
capability of line interactions to lengthen the photon propagation trajectory.
By this process, photons may potentially exert a stronger acceleration onto the
wind material, compared to the unattenuated case in which they may also interact
multiple times but their trajectories are always straight lines. Once the
unattenuation of the radiation is dropped in the \mcrh simulations, some packets
may potentially be back-scattered onto the stellar disc. To counteract the
luminosity loss induced by this process, we follow \citet{Lucy1993} and scale
the packet energies by a constant factor, which ensures that in each time step,
net radiative energy amounting to the luminosity $L_{\star}$ is streaming into
the wind. This process may be interpreted as a colour correction of the stellar
spectral energy distribution \citep[see][for a similar strategy in the context
  of calculating synthetic observables for supernovae]{Lucy1999a}. 

\begin{figure}
  \begin{center}
    \includegraphics[]{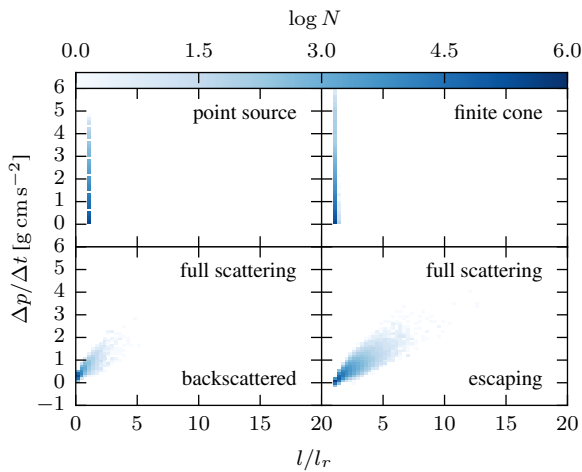}
  \end{center}
  \caption{Illustration of the total radiation momentum transfer rate exerted
    by the Monte Carlo packets along their propagation paths in terms of a
    two-dimensional histogram. The length of the trajectories, $l$, is
    normalised to the radial extent of the domain, $l_r$. The upper two panels
    show the accumulated momentum in the point-source (left) and the finite-cone
    (right) \mcrh calculation respectively. The lower panels correspond to the
    \mcrh wind simulation with the full scattering procedure. The left panel
    only includes packets that are ultimately backscattered onto the central
    star. Thus, trajectories shorter than $l_r$ are encountered. In the lower
    right panel only packets which ultimately escape through the outer boundary
    of the computational domain are shown.}
  \label{fig:wind_acc_gline_cmp}
\end{figure}

Accounting both for the finite-cone effect and the full scattering procedure, we
again determine the stationary wind structure with \mcrh.  The resulting wind
velocity, the density stratification, and the mass-loss rate are included in the
summary plot of Figure \ref{fig:wind_cmp}. Compared to the unattenuated
calculations with the finite-cone effect, we find a slightly slower wind and, as
shown in Figure \ref{fig:wind_line_cmp}, a weaker radiative acceleration. At
first glance, this seems to contradict the statement about the effect of
multiple scattering in the introductory part of this section. But one has to
bear in mind that in the \cak/\mcak description of the line-driving problem, in
each interaction the full photon/packet momentum is transferred onto the wind
material. This is comparable to a purely absorbing medium, with the important
difference that the photon trajectory is not terminated in the \cak/\mcak
description. Instead, the same photon may still contribute to the acceleration
in regions of the wind located at larger radii. In the full scattering case, in
which the re-emission of the line-interaction is taken into account, no momentum
is transferred onto the wind on a \cak/\mcak like trajectory since it involves
forward-scatterings only. Thus, the momentum transfer in the \cak/\mcak case may
generally be overestimated and only the photons that are on trajectories that
have been significantly lengthened due to many successive scatterings may
contribute comparably to the \cak/\mcak description. This phenomenon is
illustrated in Figure \ref{fig:wind_acc_gline_cmp}. A general reduction of the
line-driving force once the unattenuation assumption is dropped has already been
found in previous studies, e.g. \citet{Abbott1985}.

\subsubsection{Additional diagnostics}
\label{sec:diagnostics}

\begin{figure*}
  \begin{center}
    \includegraphics[]{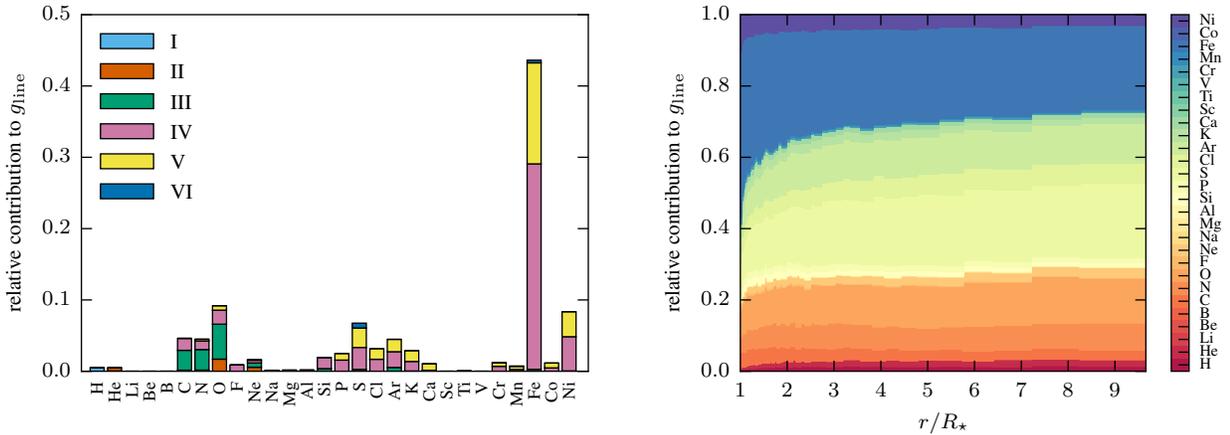}
  \end{center}
  \caption{Detailed summary of the contribution
    of the different elements to the line-driving force in the full-scattering
    simulation. In the left panel, the composition of the line-driving force
    averaged over the entire simulation box is displayed. Here, also the
    contributions of the different ionisation stages are shown. The right panel shows
    how this composition varies throughout the wind. For this illustration, the
    contributions to the line-driving for of all ions of one element are
    combined and then colour-coded. Moreover, a binning with a width of ten grid
    cells has been used.}
  \label{fig:line_force_composition}
\end{figure*}

By recording the details of all interaction events performed by the Monte Carlo
packets in the final calculation of the simulation series, the origin of the
radiative acceleration can be studied. The contributions of the different
elements and ionisation stages, averaged over the entire computational domain,
are illustrated in Figure
\ref{fig:line_force_composition}.
This highlights that the line-driving force mainly derives from interactions with
lines of iron, nickel some intermediate mass elements and the CNO group. The
relative importance of the different contributions, however, changes throughout
the wind, as shown in the right panel of the same figure. In our simulation,
lines of iron group elements mostly contribute in the inner part of the wind,
close to the photosphere. Further out, the intermediate mass elements grow in
importance. This finding is compatible with the investigation of
\citet{Vink1999}, in which the importance of iron for the radiative acceleration
in the lower parts of the wind has been highlighted in the context of the
bi-stability jump. We stress, however, that due to the simplified treatment of
ionisation and excitation in our simulations, the results presented in
Figure \ref{fig:line_force_composition} should not be over-interpreted, but
viewed as an illustration of the diagnostic possibilities of our approach. 

\begin{figure}
  \begin{center}
    \includegraphics[]{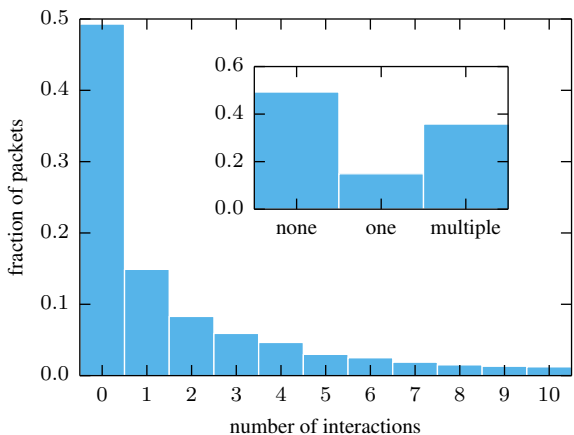}
  \end{center}
  \caption{Histogram of the number of
    interactions which Monte Carlo packets that ultimately escape through the outer
    edge of the simulation box perform. As displayed in the inset, the majority
    of the escaping packets never experience an interaction. However, most of the
    interacting packets perform multiple scatterings.}
  \label{fig:mult_interaction_hist}
\end{figure}

By recording the interaction histories of all packets we can also investigate
the importance of multiple scattering in our simulation.  Figure
\ref{fig:mult_interaction_hist} shows the number of interactions performed by
the packets that escaped through the outer edge of the computational domain. 
In our particular simulation, most packets never explicitly interacted on their
way out. They may have still contributed to the line-driving force as long as
they came into resonance with at least one line transition (see discussion in
Section \ref{sec:estimators}).  The majority of those that interacted, however,
performed multiple scattering events. 

\section{Comparison with detailed calculations}
\label{sec:mv08_sims}

Having established the basic performance of our method under simplified
conditions, we now compare our approach with modern, sophisticated
techniques for determining the structure of hot-star winds, in particular with
the approach developed by \citet[][MV08 hereafter]{Mueller2008}. This method
constitutes an advancement of the original approach of \citet{Vink1999} and
\citet{Vink2000}. A Monte Carlo radiative transfer calculation is used to
determine the local line-driving force for a given static wind structure. Here a
velocity structure that is more general and flexible than the $\beta$-type law
is used. According to the reconstructed line acceleration, the mass-loss rate
and the parameters of the wind velocity law are iteratively updated until a
converged wind structure has been found.

\subsection{Parameters}
\label{sec:mv08_pars}

MV08 test their approach by predicting the wind structure of an O5-V main
sequence star. For direct comparison, we adopt parameters in our simulation to
match their calculation. In Table \ref{tab:o5v_pars}, all quantities which have
been changed with respect to the calculations presented in Section
\ref{sec:wind_sims} are listed.
\begin{table}
  \centering
  \caption{Simulation parameters for the calculation of the
    wind from the O5-V main sequence star
    \citep[c.f.][]{Mueller2008}.}
  \begin{tabular}{cc}
    \hline
    Parameter & Value\\\hline
    $M_{\star}$ & $40\, \mathrm{M}_{\odot}$\\
    $\log L_{\star}/\mathrm{L}_{\odot}$ & $5.5$\\
    $T_{\mathrm{eff}}$ & $\SI{4e4}{K}$\\
    $N_{\mathrm{e}}/W$ & $\SI{e14}{cm^{-3}}$\\
    $\rho_0$ & $\SI{e-12}{g.cm^{-3}}$\\
    \hline
  \end{tabular}
  \label{tab:o5v_pars}
\end{table}
These choices imply an Eddington factor of $\Gamma_e = 0.210$,
which is very close to the value quoted by MV08.

The comparison with the \citet{Mueller2008} technique will be based on \mcrh
calculations that incorporate all techniques described in Section
\ref{sec:numerics}. Since no \cak or \mcak-like simulations are performed in
this context, the \cak fitting procedure of Section \ref{sec:fitting} does not
have to be repeated for the current stellar parameters.

\subsection{Results}
\label{sec:mv08_results}

\begin{figure}
  \begin{center}
    \includegraphics[]{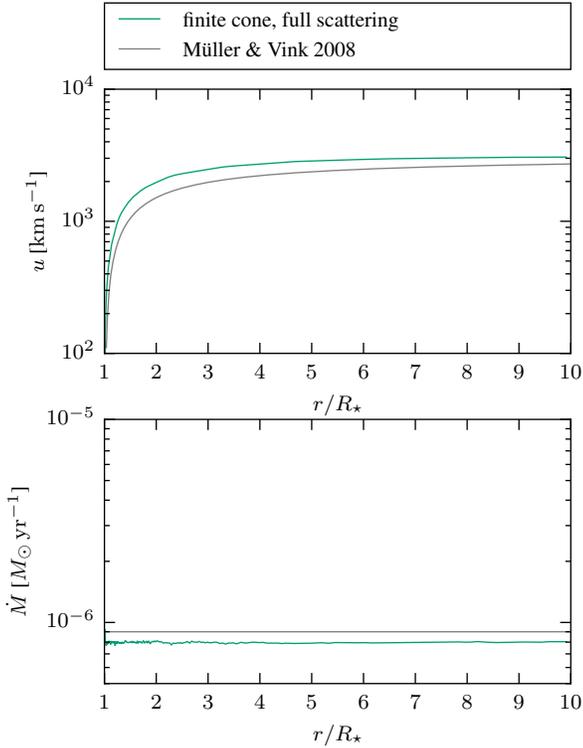}
  \end{center}
  \caption{Stationary wind structure for the O5-V main sequence star determined
    with \mcrh (green). The predictions according to the MV08
  technique are shown as a comparison (grey).}
  \label{fig:mv08cmp}
\end{figure}

With the stellar parameters listed in Table \ref{tab:o5v_pars} a full scattering
\mcrh simulation, similar to the calculation presented in Section
\ref{sec:res_att}, is performed to determine the structure of the wind of the
O5-V main sequence star. The result is shown in Figure \ref{fig:mv08cmp} in
terms of the stationary velocity and mass-loss rate and compared to the
structure found by MV08.
To obtain the comparison MV08 wind velocity, we use the approximate expression,
Equation 39 in MV08, which is only strictly valid in the supersonic wind regime. 

As seen in Figure \ref{fig:mv08cmp}, both approaches predict winds that are
quite similar in shape. However, the velocity structure found by \mcrh
accelerates quite quickly towards the terminal wind speed, whereas the wind
velocity law predicted by MV08 approaches its final value in a gentler fashion.
The two winds deviate slightly in absolute values: the \mcrh wind reaches a
velocity of $u = \SI{3065}{km.s^{-1}}$ at $r = 10\, R_{\star}$, while the MV08
solution lies at $u = \SI{2719}{km.s^{-1}}$. The mass-loss rates of the winds
agree on a similar, namely a ten percent level, with $\dot M = 7.97 \times
10^{-7} \, M_{\odot}\,\mathrm{yr}^{-1}$ being obtained in the \mcrh calculation
and $\dot M = 8.99 \times 10^{-7} \, M_{\odot}\,\mathrm{yr}^{-1}$ found by MV08.

These minor discrepancies may partly be related to the different solution
strategies followed in the two approaches. In our method, a radiation
hydrodynamical calculation is relaxed to a steady state solution without any
major restrictions on the shape of the line acceleration and velocity structure.
The MV08 approach relies on a parameterisation of the line driving force and in
turn of the velocity law, thus allowing only wind structures of a certain
family.  Differences between the two calculations will also arise because of the
very simplistic treatment of ionisation and excitation in our method. Overall,
however, the level of agreement between the two calculations is good and lends
credence to their complementary use in the study of stellar winds.

\section{Discussion}
\label{sec:discussion}

The results of the wind calculations and the diagnostic capabilities presented
in the previous sections, in particular the good agreement between the \mcrh
results and the \cak/\mcak predictions and the comparison to the MV08 technique,
clearly demonstrate the utility and accuracy of our Monte Carlo-based radiation
hydrodynamical approach for solving the line-driving problem in hot-star winds.
In the following, we briefly discuss some particular features to our approach
and also comment on some of its current limitations. We also sketch possible
alterations of our approach aimed at alleviating these shortcomings and discuss
potential future applications.

\subsection{Numerical performance}

In general, Monte Carlo radiative transfer techniques are rather computationally
demanding, since many Monte Carlo quanta have to be processed to reach the
desired statistical fidelity. This drawback is typically balanced by the very
favourable parallelisation properties of such calculations (see discussion in
section \ref{sec:nonsob}). In our case, the Monte Carlo-based radiation
hydrodynamical simulations are indeed costly. In the full scattering simulation,
the final stationary state established after about 4800 simulation cycles. On of
these takes roughly $\SI{14}{s}$ on a Intel Xeon E5520 processor if
$5\times10^4$ packets describe the incident radiation field. This increases to
about $\SI{210}{s}$ once $10^6$ packets are used. In these calculations about
$10\,\mathrm{GFlops}$ (for $5 \times 10^4$ packets) and $160\, \mathrm{GFlops}$
($10^6$ packets) were executed per cycle. By comparison, only $\SI{0.015}{s}$ are required
and $1.3\, \mathrm{MFlops}$ executed per cycle if the \cakrh version is used instead. We
emphasise, however, that all calculations presented in this work were carried
out serially on a single processing core.  We stress, that the performance of
the Monte Carlo scheme may be significantly improved by devising a
parallelisation scheme and distributing the workload onto many cores. For all
calculations, the executable was produced with the \texttt{gcc} compiler,
version 4.7.3, using the optimisation level \texttt{-O3}.

\subsection{Influence of Monte Carlo noise} \label{sec:mc_noise}

The probabilistic nature of Monte Carlo techniques leads to the inevitable
introduction of stochastic fluctuations (see, for example, \citet{Carter1975}
for a discussion of Monte Carlo errors). However, \citet{Noebauer2012} and
\citet{Roth2015} demonstrated that, with appropriate reconstruction techniques,
the fluctuations can be controlled and Monte Carlo-based radiation
hydrodynamical simulations can indeed be performed. The simulations performed in
this work and presented in the previous section illustrate that the Monte Carlo
noise, which is clearly visible in the reconstructed radiative acceleration
(c.f.\  Figure \ref{fig:wind_line_cmp}), does not prohibit the use of Monte
Carlo methods to study the line-driving problem. In fact, the obtained velocity
and density structure (shown in Figure \ref{fig:wind_cmp}) is remarkably smooth.
Due to their pure stochastic nature the fluctuations cancel in an average sense
(but see also remarks in Section \ref{sec:nonsob}).

\subsection{Full non-LTE treatment}
\label{sec:nonlte}

The calculations presented here assumed either LTE or relied on the approximate
non-LTE prescription presented in Section \ref{sec:ionisation}. Non-LTE effects,
however, play an important role in stellar winds \citep[see, e.g.,][for non-LTE
calculations of hot-star winds]{Pauldrach1987,Pauldrach1994,Puls2005} and should
be included in future calculations.

In the general non-LTE situation, the level populations of excited states are
influenced by the radiation field. \citet{Lucy2002} and \citet{Lucy2003}
describes a simple and elegant procedure to reconstruct these radiative rates
from a Monte Carlo simulation. In a first step of refining the physical detail
in the interaction treatment of our approach, this procedure could be adopted to
determine a non-LTE excitation and ionisation balance during the Monte Carlo
radiative transfer step. This strategy has already been followed in previous
Monte Carlo-based studies, for example by \citet{Sim2005a}, using the
\textsc{Python} code \citep{Long2002} and in non-LTE radiative transfer
calculations in Keplerian discs by \citet{Carciofi2006}.

To go beyond the pure resonant scattering approximation in the line-interaction
treatment, either the simple branching scheme of \citet{Lucy1999a} (see
\citealt{Mazzali2000}, for Monte Carlo-based radiative transfer calculations in
Type Ia ejecta using this scheme) may be incorporated or the so-called
\textit{Macro-Atom} formalism introduced by \citet{Lucy2002,Lucy2003} may be
used. The latter fully accounts for all processes that may excite or de-excite
line-transitions, including collisional processes. We emphasise, that the
utility and accuracy of these techniques has been carefully examined and
established \citep{Lucy2002,Lucy2003,Lucy2005}. Moreover, they have already been
incorporated into a number of Monte Carlo radiative transfer frameworks,
including \textsc{Artis} \citep{Kromer2009}, \textsc{Python} \citep{Sim2005a}
and \textsc{Tardis} \citep{Kerzendorf2014}.

\subsection{Multidimensionality and non-Sobolev treatment}
\label{sec:nonsob}

Observations and also theoretical investigations indicate that
line-driven winds are neither stationary nor smooth, but feature a variable and
inhomogeneous structure \citep[see, e.g., overview by][]{Puls2008}. These findings advocate a dynamical and fully
multidimensional treatment of the problem. Monte Carlo techniques are 
suited for this task since the formalism readily generalises (conceptually) to
arbitrary geometrical configurations and also exhibits excellent scaling
properties on parallel processors -- a desirable feature when facing the
increased numerical workload of multidimensional calculations (see, for example,
comments by \citealt{Kasen2006} and \citealt{Baes2011}, as well as the scaling
tests by \citealt{Roth2015} and \citealt{Harries2015}). With this in mind an obvious extension of our method comprises a
generalisation to multidimensional geometries to address inhomogeneous
line-driven outflows. In this context, some work should be invested into
devising an efficient interfacing of the standard parallelisation strategies of
Monte Carlo and fluid dynamical techniques (see e.g.\ discussion in
\citealt{Baes2011} and developments by \citealt{Harries2015}.).

When investigating multidimensional line-driven mass outflows, a generalisation
of the line interaction treatment that goes beyond the Sobolev approximation
used here should also be considered: this would be necessary to study the line-driving
instability \citep{Lucy1970,Owocki1994} which should occur in line-driven
outflows and may play a part in understanding the clumping mechanism (see
discussion in \citealt{Puls2008} and \citealt{Vink2015}). Non-Sobolev Monte
Carlo radiative transfer schemes have already been developed and used, for
example by \citet{Knigge1995} and \citet{Kusterer2014} in the context of winds
of cataclysmic variables.  However, the computational costs of such schemes are
significantly higher than their Sobolev counterparts. Moreover, the inherent
Monte Carlo noise could potentially be more problematic than described in
\ref{sec:mc_noise}, since small perturbations in the line-driven wind may, in
principle, self amplify in non-Sobolev calculations. Further investigation is
required to asses the relevance of this potential caveat.

\section{Conclusions}

In this work we have introduced a new approach to solve line-driven stellar
winds self-consistently by using a Monte Carlo-based radiation hydrodynamical
approach. The key feature of this technique lies in the reliance on a 
Monte Carlo radiative scheme. This technique offers a number of advantages when
dealing with complex interaction physics, in particular when multi-line effects
play a role. Moreover, this Monte Carlo-based technique readily generalises to
multidimensional geometries, a very advantageous feature for potential studies
of inhomogeneous outflows.

Establishing the utility and accuracy of the introduced approach in solving the
line-driving problem was the main focus of this work.  Consequently, we designed
the calculations to capture the essence of the line-driving problem, but adopted
a number of simplifications. These do not interfere with the general
applicability of our method to solve the local line-driving problem, but reduce
the computational complexity. Most importantly, we adopted the Sobolev
approximation and a simple and approximate non-LTE treatment to determine the
excitation and ionisation balance.

Using our scheme, we successfully solved for the stationary structure of
one-dimensional, spherically symmetric hot-star winds achieving good agreement
with the predictions of the \cak and \mcak theory. We demonstrated that our
method can also go beyond the capabilities of \cak and \mcak by dealing with an
attenuated radiation field and the effects of multiple scattering. For the
particular physical conditions investigated in Section \ref{sec:wind_sims},
these effects lead to a reduction of the line-driving force and thus a slower
wind velocity structure compared to the \mcak predictions. Finally, we compared
results obtained with our approach to those computed by MV08 for an O5-V main
sequence star and found good agreement, given the difference in approach and
simplifications made here.

The successful outcome of our wind simulations demonstrates the possibility to
use a Monte Carlo-based radiation hydrodynamical approach to model line-driven
mass outflows. Thus, this approach holds promise for detailed multidimensional
and self-consistent studies of inhomogeneous winds, including multi-line
effects. With a fully multidimensional version of our approach, line-driven
outflows in systems other than hot-star winds may be addressed as well. In
particular, the winds emanating from accretion discs in cataclysmic variables
\citep[e.g.][]{Proga1998,Noebauer2010} or active galactic nuclei
\citep[e.g.][]{Proga2000, Proga2004, Higginbottom2013} may be investigated
self-consistently and in great detail with such a Monte Carlo-based approach.

\section*{Acknowledgements}

UMN owes his gratitude to Wolfgang Hillebrandt for his support and supervision
along the PhD project, during which part of the work presented here has been
performed. The authors thank Jorick Vink for his valuable comments on the
manuscript. Markus Kromer, Philipp Edelmann, Rolf-Peter Kudritzki, Friedrich
R{\"o}pke and Ewald M{\"u}ller are thanked for many fruitful discussions.  The
authors also thank the anonymous reviewer for valuable comments which greatly
improved the paper. UMN is supported by the Transregional Collaborative Research
Centre TRR 33 ``The Dark Universe'' of the German Research Foundation (DFG).
Figures were produced using \texttt{matplotlib} \citep{Hunter2007}. 

\appendix

\section{CAK and MCAK Approaches}
\label{app:cak}

In the following, the relevant expressions of the \cak theory are listed in
the form in which they are used in the current study. In general, the
line-driving force is expressed as a multiple of the acceleration due to
electron scattering
\begin{equation}
  g_{\mathrm{line}} = M(t) \frac{\sigma_{\mathrm{e}}^{\mathrm{ref}} L_{\star}}{4
  \pi r^2 c},
  \label{eq:app_cak_line}
\end{equation}
assuming a reference specific interaction cross section
$\sigma^{\mathrm{ref}}_{\mathrm{e}}$. Throughout this work,
$\sigma_{\mathrm{e}}^{\mathrm{ref}} = \SI{0.3}{cm^2.g^{-1}}$ is used.
The force multiplier depends on the dimensionless optical-depth $t$
\begin{equation}
  t = \sigma_{\mathrm{e}}^{\mathrm{ref}} \rho u_{\mathrm{th}}
  \left(\frac{\mathrm{d}u}{\mathrm{d}r}\right)^{-1},
  \label{eq:app_cak_t}
\end{equation}
which involves the thermal velocity of the wind material \citep[c.f.][]{Abbott1982}
\begin{equation}
  u_{\mathrm{th}} = \sqrt{\frac{2 k_{\mathrm{B}}T_{\mathrm{eff}}}{m_{\mathrm{H}}}},
  \label{eq:app_cak_vth}
\end{equation}
and is approximated by the power-law
\begin{equation}
  M(t) = k t^{-\alpha}
  \label{eq:app_cak_mt}
\end{equation}
if the central star is assumed to radiate as a point-source (again
following \citealt{Castor1975}). It should be noted that, throughout this work,
we neglect any influence of changes in the ionisation, which is equivalent to
setting $\delta = 0$ in the force multiplier formulation proposed by
\citet{Abbott1982}.

In the point-source limit, the solution to the momentum equation
(\ref{eq:force_balance_wind}), with the \cak line-driving force, results in the
wind velocity law
\begin{equation}
  u(r) = u_{\infty} \sqrt{1 - \frac{R_{\star}}{r}},
  \label{eq:app_cak_velocity_law}
\end{equation}
with the terminal speed being a multiple of the local escape speed from the
photosphere, $u_{\mathrm{esc}}$:
\begin{equation}
  u_{\infty} = \sqrt{\frac{\alpha}{1 - \alpha}} u_{\mathrm{esc}}.
  \label{eq:app_cak_terminal_wind_speed}
\end{equation}
The constant mass-loss rate of the wind is given by
\begin{align}
  \dot M_{\mathrm{CAK}} &= \frac{1}{u_{\mathrm{th}}} \left(
  \frac{4\pi}{\sigma_{\mathrm{e}^{\mathrm{ref}}}} G M_{\star} (1 -
  \Gamma_{\mathrm{e}} \right)^{1 - \frac{1}{\alpha}} \left( \frac{k}{c}L_{\star}
  \right)^{\frac{1}{\alpha}}\nonumber\\
  &\times \alpha (1 - \alpha)^{\frac{1}{\alpha} - 1}.
  \label{eq:app_cak_mass_loss_rate}
\end{align}

Once the finite extent of the star is taken into account, the force multiplier
is modified by a finite-cone correction factor
\begin{align}
  M_{\mathrm{FC}}(t) &= D_{\mathrm{FC}}(t) M_{\mathrm{CAK}}(t),
  \label{eq:app_mcak_mt}\\
  D_{\mathrm{FC}}(t) & = \frac{(1 + \sigma)^{\alpha + 1} - (1 + \sigma
  \mu_{\star}^2)^{\alpha+1}}{(1 - \mu_{\star}^2)(\alpha+1)\sigma(1 +
  \sigma)^{\alpha}},
  \label{eq:app_mcak_mt_df}
\end{align}
which involves 
\begin{align}
  \sigma &= \frac{r}{u} \frac{\mathrm{d}u}{\mathrm{d}r} - 1,
  \label{eq:app_sigma}\\
  \mu_{\star} & = \sqrt{1 - \left(\frac{R_{\star}}{r}\right)^2}.
  \label{eq:app_limiting_mu}
\end{align}
We follow \citet{Kudritzki1989} and predict the wind structure according to
their approximate analytic solution technique. In particular, we adopt
their proposed approach for the case of a ``frozen-in'' ionisation state
($\delta = 0$, c.f.\ \citealt{Kudritzki1989}, section 4.1). The wind velocity is assumed to be very close
to a $\beta$-type law and the mass-loss rate decreases
with respect to the original \cak case according to
\begin{equation}
  \dot M = \left( \frac{1}{1 + \alpha} \right)^{\frac{1}{\alpha}}\dot
  M_{\mathrm{CAK}}.
  \label{eq:mcak_mdot}
\end{equation}
The wind velocity in this approach follows from performing the integration
\begin{equation}
  u(v) = \sqrt{\frac{\alpha}{1 - \alpha} u_{\mathrm{esc}}^2 \int_v^1 \mathrm{d}
  v' Z(v', \alpha, \beta)},
  \label{eq:mcak_velocity_integral}
\end{equation}
with
\begin{align}
  Z(v, \alpha, \beta) &= f_N(v, \alpha, \beta)^{\frac{1}{1-\alpha}}
  \nonumber\\&\times\left[
  1 + \sqrt{\frac{2}{\alpha}\left( 1 - \left(\frac{1}{f_N(v, \alpha,
  \beta)}\right)^{\frac{1}{1 - \alpha}} \right)} \right],
  \label{eq:mcak_velocity_integral_integrand}\\
  f_N(v, \alpha, \beta) &= \frac{\beta}{v} \frac{1}{v(\beta + 1) - 1}
   \nonumber\\&\times \left[ 1 - \left( 1 - v^2 + v\frac{1 - v}{\beta} 
  \right)^{\alpha+1} \right].
  \label{eq:mcak_velocity_integral_fN}
\end{align}
In these expressions, the inverse of the radial distance relative to the
photosphere, $v = R_{\star}/r$, is used. Throughout this work, the \mcak
equations are solved for $\beta = 0.8$ \citep[see][for a motivation of this
value]{Pauldrach1986,Kudritzki1989}.

\section{Radiation Hydrodynamics Equations}
\label{app:rh_equations}

We briefly review the radiation hydrodynamical equations on which our numerical
scheme is based \citep[c.f.][]{Mihalas1984}. In general, these equations
describe the conservation of mass, momentum and energy. Since we assume in this
work, that the wind outflow remains isothermal at all times, we only consider
the first two of these equations and use the isothermal equation of state to
relate fluid density and thermodynamic pressure. Cast into a pseudo-Lagrangian
form by using the substantial derivative
\begin{equation}
  \frac{\mathrm{D}}{\mathrm{D} t} = \frac{\mathrm{d}}{\mathrm{d} t} + u
  \frac{\mathrm{d}}{\mathrm{d} r},
  \label{eq:substantial_derivative}
\end{equation}
this radiation hydrodynamical problem in one-dimensional spherical symmetry is
described by \citep[c.f.][]{Mihalas1984}
\begin{align}
  \frac{\mathrm{D}}{\mathrm{D} t} \rho + \rho \frac{1}{r^2}
  \frac{\mathrm{d}}{\mathrm{d} r} (r^2 u) &= 0,
  \label{eq:radiation_hydrodynamics_mass}\\
  \rho \frac{\mathrm{D}}{\mathrm{D} t} u + \frac{\mathrm{d}}{\mathrm{d} r} P & =
  \rho g +
  G^1,
  \label{eq:radiation_hydrodynamics_momentum}\\
  P &= a_{\mathrm{iso}}^2 \rho.
  \label{eq:radiation_hydrodynamics_eos}
\end{align}
The fluid density $\rho$ and its velocity $u$ appear together with the
thermodynamic pressure $P$, the isothermal sound speed $a_{\mathrm{iso}}$ and a static external
gravitational field $g$. The transfer of momentum between the fluid and the
radiation field is captured by the radiation force, which may be determined from
the first moment of the transfer equation \citep[c.f.][]{Mihalas1984}:
\begin{equation}
  G^1 = \frac{2 \pi}{c} \int_0^{\infty} \mathrm{d} \nu \int_{-1}^{1}
  \mathrm{d} \mu (\chi I - \eta) \mu.
  \label{eq:radiation_force_1}
\end{equation}
Here, the description of the radiation field by the specific intensity $I$ is
used and the material functions opacity $\chi$ and emissivity $\eta$, describing
the absorption and emission of radiative energy, appear. The integration is
performed with respect to the entire frequency spectrum and to all possible
values for the cosine of the propagation direction, $\mu$.

%%\bibliography{astrouli.bib}
%%\bibliographystyle{mn2e}

\end{document}

%% file: velocity_interp_cell.tikz
\begin{tikzpicture}[scale=1.75]

\def\xarray{{4.2,4.7,5.2,5.7,6.2}};
\def\yarray{{0.872871560944,1.54368996998,1.92153784566,2.18447472509,2.38273358873}};
\def\xlarray{{3.95,4.45,4.95,5.45,5.95}};
\def\xrarray{{4.45,4.95,5.45,5.95,6.45}};
\def\yiarray{{1.20828076546,1.73261390782,2.05300628538,2.28360415691}};
%\definecolor{darkmint}{rgb}{0.4862745,0.6509803,0.6509803}%
\definecolor{skyblue}{rgb}{0.33725490196078434, 0.7058823529411765, 0.9137254901960784}

\draw[arrows=-stealth] (\xlarray[0] - 0.25, 0) -- (\xlarray[0]- 0.25, \yarray[4] + 0.5);

\node[anchor = south, rotate = 90] at (\xlarray[0]- 0.25, {0.5 * (\yarray[4] + 0.5)}) {$u$};

\node[anchor=north] at (\xarray[0],0) {$i-2$};
\node[anchor=north] at (\xarray[1],0) {$i-1$};
\node[anchor=north] at (\xarray[2],0) {$i$};
\node[anchor=north] at (\xarray[3],0) {$i+1$};
\node[anchor=north] at (\xarray[4],0) {$i+2$};

\foreach \i in {0,...,4}
{
	\pgfmathsetmacro{\x}{\xarray[int(\i)]};
	\pgfmathsetmacro{\y}{\yarray[int(\i)]};
	\filldraw[fill=black,draw=black] (\x, \y]) circle (0.75pt);
	\draw (\xlarray[\i], 0) rectangle (\xrarray[\i], \y);
}
\foreach \i in {0,...,3}
{
	\draw[dashed] (\xarray[\i], \yarray[\i]) -- (\xarray[\i+1], \yarray[\i+1]);
	\filldraw[fill=skyblue,draw=skyblue] (\xrarray[\i], \yiarray[\i]) circle (0.75pt);
}

\foreach \i in {0,...,2}
{
	\draw[color=skyblue] (\xrarray[\i], \yiarray[\i]) -- (\xrarray[\i+1], \yiarray[\i+1]);
}

\end{tikzpicture}

%% file: noebauer_sim.bbl
\begin{thebibliography}{75}
\expandafter\ifx\csname natexlab\endcsname\relax\def\natexlab#1{#1}\fi

\bibitem[{{Abbott}(1982)}]{Abbott1982}
{Abbott} D.~C., 1982, \apj, 259, 282

\bibitem[{{Abbott} \& {Lucy}(1985)}]{Abbott1985}
{Abbott} D.~C., {Lucy} L.~B., 1985, \apj, 288, 679

\bibitem[{{Acreman} {et~al}\mbox{.}(2010){Acreman}, {Harries}, \&
  {Rundle}}]{Acreman2010}
{Acreman} D.~M., {Harries} T.~J., {Rundle} D.~A., 2010, \mnras, 403, 1143

\bibitem[{{Asplund} {et~al}\mbox{.}(2009){Asplund}, {Grevesse}, {Sauval}, \&
  {Scott}}]{Asplund2009}
{Asplund} M., {Grevesse} N., {Sauval} A.~J., {Scott} P., 2009, \araa, 47, 481

\bibitem[{{Baes} {et~al}\mbox{.}(2011){Baes}, {Verstappen}, {De Looze},
  {Fritz}, {Saftly}, {Vidal P{\'e}rez}, {Stalevski}, \& {Valcke}}]{Baes2011}
{Baes} M., {Verstappen} J., {De Looze} I., {Fritz} J., {Saftly} W., {Vidal
  P{\'e}rez} E., {Stalevski} M., {Valcke} S., 2011, \apjs, 196, 22

\bibitem[{Batten {et~al}\mbox{.}(1997)Batten, Clarke, Lambert, \&
  Causon}]{Batten1997}
Batten P., Clarke N., Lambert C., Causon D., 1997, SIAM Journal on Scientific
  Computing, 18, 1553

\bibitem[{{Camps} {et~al}\mbox{.}(2013){Camps}, {Baes}, \&
  {Saftly}}]{Camps2013}
{Camps} P., {Baes} M., {Saftly} W., 2013, \aap, 560, A35

\bibitem[{{Carciofi} \& {Bjorkman}(2006)}]{Carciofi2006}
{Carciofi} A.~C., {Bjorkman} J.~E., 2006, \apj, 639, 1081

\bibitem[{{Carter} \& {Cashwell}(1975)}]{Carter1975}
{Carter} L.~L., {Cashwell} E., 1975, {Particle-transport simulation with the
  Monte Carlo method}. Tech. rep.

\bibitem[{{Castor} {et~al}\mbox{.}(1975){Castor}, {Abbott}, \&
  {Klein}}]{Castor1975}
{Castor} J.~I., {Abbott} D.~C., {Klein} R.~I., 1975, \apj, 195, 157

\bibitem[{{Colella} \& {Woodward}(1984)}]{Colella1984}
{Colella} P., {Woodward} P.~R., 1984, Journal of Computational Physics, 54, 174

\bibitem[{{Dessart}(2004)}]{Dessart2004}
{Dessart} L., 2004, \aap, 423, 693

\bibitem[{{Dessart} \& {Owocki}(2003)}]{Dessart2003}
{Dessart} L., {Owocki} S.~P., 2003, \aap, 406, L1

\bibitem[{{Dessart} \& {Owocki}(2005)}]{Dessart2005}
{Dessart} L., {Owocki} S.~P., 2005, \aap, 437, 657

\bibitem[{{Feldmeier}(1995)}]{Feldmeier1995}
{Feldmeier} A., 1995, \aap, 299, 523

\bibitem[{{Feldmeier} {et~al}\mbox{.}(1997){Feldmeier}, {Kudritzki}, {Palsa},
  {Pauldrach}, \& {Puls}}]{Feldmeier1997}
{Feldmeier} A., {Kudritzki} R.-P., {Palsa} R., {Pauldrach} A.~W.~A., {Puls} J.,
  1997, \aap, 320, 899

\bibitem[{Fleck \& Cummings(1971)}]{Fleck1971}
Fleck J.~A., Cummings J.~D., 1971, Journal of Computational Physics, 8, 313

\bibitem[{Fornberg(1988)}]{Fornberg1988}
Fornberg B., 1988, Mathematics of Computation, 51, pp. 699

\bibitem[{{Friend} \& {Abbott}(1986)}]{Friend1986}
{Friend} D.~B., {Abbott} D.~C., 1986, \apj, 311, 701

\bibitem[{{Hamann} \& {Koesterke}(1998)}]{Hamann1998}
{Hamann} W.-R., {Koesterke} L., 1998, \aap, 335, 1003

\bibitem[{{Harries}(2015)}]{Harries2015}
{Harries} T.~J., 2015, \mnras, 448, 3156

\bibitem[{{Haworth} \& {Harries}(2012)}]{Haworth2012}
{Haworth} T.~J., {Harries} T.~J., 2012, \mnras, 420, 562

\bibitem[{{Higginbottom} {et~al}\mbox{.}(2013){Higginbottom}, {Knigge}, {Long},
  {Sim}, \& {Matthews}}]{Higginbottom2013}
{Higginbottom} N., {Knigge} C., {Long} K.~S., {Sim} S.~A., {Matthews} J.~H.,
  2013, \mnras, 436, 1390

\bibitem[{{Hunter}(2007)}]{Hunter2007}
{Hunter} J.~D., 2007, Computing in Science and Engineering, 9, 90

\bibitem[{{Kasen} {et~al}\mbox{.}(2006){Kasen}, {Thomas}, \&
  {Nugent}}]{Kasen2006}
{Kasen} D., {Thomas} R.~C., {Nugent} P., 2006, \apj, 651, 366

\bibitem[{{Kerzendorf} \& {Sim}(2014)}]{Kerzendorf2014}
{Kerzendorf} W.~E., {Sim} S.~A., 2014, \mnras, 440, 387

\bibitem[{{Klein} \& {Castor}(1978)}]{Klein1978a}
{Klein} R.~I., {Castor} J.~I., 1978, \apj, 220, 902

\bibitem[{{Knigge} {et~al}\mbox{.}(1995){Knigge}, {Woods}, \&
  {Drew}}]{Knigge1995}
{Knigge} C., {Woods} J.~A., {Drew} J.~E., 1995, \mnras, 273, 225

\bibitem[{{Kromer} \& {Sim}(2009)}]{Kromer2009}
{Kromer} M., {Sim} S.~A., 2009, \mnras, 398, 1809

\bibitem[{{Kudritzki} {et~al}\mbox{.}(1989){Kudritzki}, {Pauldrach}, {Puls}, \&
  {Abbott}}]{Kudritzki1989}
{Kudritzki} R.~P., {Pauldrach} A., {Puls} J., {Abbott} D.~C., 1989, \aap, 219,
  205

\bibitem[{{Kudritzki} \& {Puls}(2000)}]{Kudritzki2000}
{Kudritzki} R.-P., {Puls} J., 2000, \araa, 38, 613

\bibitem[{{Kurucz} \& {Bell}(1995)}]{Kurucz1995}
{Kurucz} R.~L., {Bell} B., 1995, {Atomic Line List}. Kurucz CD-ROM, Cambridge,
  MA: Smithsonian Astrophysical Observatory

\bibitem[{{Kusterer} {et~al}\mbox{.}(2014){Kusterer}, {Nagel}, {Hartmann},
  {Werner}, \& {Feldmeier}}]{Kusterer2014}
{Kusterer} D.-J., {Nagel} T., {Hartmann} S., {Werner} K., {Feldmeier} A., 2014,
  \aap, 561, A14

\bibitem[{{Lamers} \& {Cassinelli}(1999)}]{Lamers1999}
{Lamers} H.~J.~G.~L.~M., {Cassinelli} J.~P., 1999, {Introduction to Stellar
  Winds}. Cambridge, UK: Cambridge University Press

\bibitem[{{Long} \& {Knigge}(2002)}]{Long2002}
{Long} K.~S., {Knigge} C., 2002, \apj, 579, 725

\bibitem[{{Lucy}(1999{\natexlab{a}})}]{Lucy1999}
{Lucy} L.~B., 1999{\natexlab{a}}, \aap, 344, 282

\bibitem[{{Lucy}(1999{\natexlab{b}})}]{Lucy1999a}
{Lucy} L.~B., 1999{\natexlab{b}}, \aap, 345, 211

\bibitem[{{Lucy}(2002)}]{Lucy2002}
{Lucy} L.~B., 2002, \aap, 384, 725

\bibitem[{{Lucy}(2003)}]{Lucy2003}
{Lucy} L.~B., 2003, \aap, 403, 261

\bibitem[{{Lucy}(2005)}]{Lucy2005}
{Lucy} L.~B., 2005, \aap, 429, 19

\bibitem[{{Lucy} \& {Abbott}(1993)}]{Lucy1993}
{Lucy} L.~B., {Abbott} D.~C., 1993, \apj, 405, 738

\bibitem[{{Lucy} \& {Solomon}(1970)}]{Lucy1970}
{Lucy} L.~B., {Solomon} P.~M., 1970, \apj, 159, 879

\bibitem[{{Mazzali}(2000)}]{Mazzali2000}
{Mazzali} P.~A., 2000, \aap, 363, 705

\bibitem[{{Mazzali} \& {Lucy}(1993)}]{Mazzali1993}
{Mazzali} P.~A., {Lucy} L.~B., 1993, \aap, 279, 447

\bibitem[{{Mihalas} \& {Mihalas}(1984)}]{Mihalas1984}
{Mihalas} D., {Mihalas} B.~W., 1984, {Foundations of Radiation Hydrodynamics}.
  New York: Oxford University Press

\bibitem[{{Muijres} {et~al}\mbox{.}(2012){Muijres}, {Vink}, {de Koter},
  {Hirschi}, {Langer}, \& {Yoon}}]{Muijres2012}
{Muijres} L., {Vink} J.~S., {de Koter} A., {Hirschi} R., {Langer} N., {Yoon}
  S.-C., 2012, \aap, 546, A42

\bibitem[{{M{\"u}ller} \& {Vink}(2008)}]{Mueller2008}
{M{\"u}ller} P.~E., {Vink} J.~S., 2008, \aap, 492, 493

\bibitem[{{Nayakshin} {et~al}\mbox{.}(2009){Nayakshin}, {Cha}, \&
  {Hobbs}}]{Nayakshin2009}
{Nayakshin} S., {Cha} S.-H., {Hobbs} A., 2009, \mnras, 397, 1314

\bibitem[{Noebauer(2014)}]{Noebauer2014}
Noebauer U.~M., 2014, Dissertation, Technische Universit{\"a}t M{\"u}nchen,
  M{\"u}nchen

\bibitem[{{Noebauer} {et~al}\mbox{.}(2010){Noebauer}, {Long}, {Sim}, \&
  {Knigge}}]{Noebauer2010}
{Noebauer} U.~M., {Long} K.~S., {Sim} S.~A., {Knigge} C., 2010, \apj, 719, 1932

\bibitem[{{Noebauer} {et~al}\mbox{.}(2012){Noebauer}, {Sim}, {Kromer},
  {R{\"o}pke}, \& {Hillebrandt}}]{Noebauer2012}
{Noebauer} U.~M., {Sim} S.~A., {Kromer} M., {R{\"o}pke} F.~K., {Hillebrandt}
  W., 2012, \mnras, 425, 1430

\bibitem[{{Owocki}(1994)}]{Owocki1994}
{Owocki} S.~P., 1994, \apss, 221, 3

\bibitem[{{Owocki} {et~al}\mbox{.}(1988){Owocki}, {Castor}, \&
  {Rybicki}}]{Owocki1988}
{Owocki} S.~P., {Castor} J.~I., {Rybicki} G.~B., 1988, \apj, 335, 914

\bibitem[{{Parkin} \& {Sim}(2013)}]{Parkin2013}
{Parkin} E.~R., {Sim} S.~A., 2013, \apj, 767, 114

\bibitem[{{Pauldrach}(1987)}]{Pauldrach1987}
{Pauldrach} A., 1987, \aap, 183, 295

\bibitem[{{Pauldrach} {et~al}\mbox{.}(1986){Pauldrach}, {Puls}, \&
  {Kudritzki}}]{Pauldrach1986}
{Pauldrach} A., {Puls} J., {Kudritzki} R.~P., 1986, \aap, 164, 86

\bibitem[{{Pauldrach} {et~al}\mbox{.}(1994){Pauldrach}, {Kudritzki}, {Puls},
  {Butler}, \& {Hunsinger}}]{Pauldrach1994}
{Pauldrach} A.~W.~A., {Kudritzki} R.~P., {Puls} J., {Butler} K., {Hunsinger}
  J., 1994, \aap, 283, 525

\bibitem[{{Pozdnyakov} {et~al}\mbox{.}(1983){Pozdnyakov}, {Sobol}, \&
  {Syunyaev}}]{Pozdnyakov1983}
{Pozdnyakov} L.~A., {Sobol} I.~M., {Syunyaev} R.~A., 1983, Astrophysics and
  Space Physics Reviews, 2, 189

\bibitem[{{Proga} \& {Kallman}(2004)}]{Proga2004}
{Proga} D., {Kallman} T.~R., 2004, \apj, 616, 688

\bibitem[{{Proga} {et~al}\mbox{.}(1998){Proga}, {Stone}, \& {Drew}}]{Proga1998}
{Proga} D., {Stone} J.~M., {Drew} J.~E., 1998, \mnras, 295, 595

\bibitem[{{Proga} {et~al}\mbox{.}(2000){Proga}, {Stone}, \&
  {Kallman}}]{Proga2000}
{Proga} D., {Stone} J.~M., {Kallman} T.~R., 2000, \apj, 543, 686

\bibitem[{{Puls} {et~al}\mbox{.}(1996){Puls}, {Kudritzki}, {Herrero},
  {Pauldrach}, {Haser}, {Lennon}, {Gabler}, {Voels}, {Vilchez}, {Wachter}, \&
  {Feldmeier}}]{Puls1996}
{Puls} J. {et~al.}, 1996, \aap, 305, 171

\bibitem[{{Puls} {et~al}\mbox{.}(2005){Puls}, {Urbaneja}, {Venero}, {Repolust},
  {Springmann}, {Jokuthy}, \& {Mokiem}}]{Puls2005}
{Puls} J., {Urbaneja} M.~A., {Venero} R., {Repolust} T., {Springmann} U.,
  {Jokuthy} A., {Mokiem} M.~R., 2005, \aap, 435, 669

\bibitem[{Puls {et~al}\mbox{.}(2008)Puls, Vink, \& Najarro}]{Puls2008}
Puls J., Vink J., Najarro F., 2008, The Astronomy and Astrophysics Review, 16,
  209

\bibitem[{{Roth} \& {Kasen}(2015)}]{Roth2015}
{Roth} N., {Kasen} D., 2015, \apjs, 217, 9

\bibitem[{{Schmutz}(1997)}]{Schmutz1997}
{Schmutz} W., 1997, \aap, 321, 268

\bibitem[{{Sim}(2004)}]{Sim2004}
{Sim} S.~A., 2004, \mnras, 349, 899

\bibitem[{{Sim} {et~al}\mbox{.}(2005){Sim}, {Drew}, \& {Long}}]{Sim2005a}
{Sim} S.~A., {Drew} J.~E., {Long} K.~S., 2005, \mnras, 363, 615

\bibitem[{{Sobolev}(1960)}]{Sobolev1960}
{Sobolev} V.~V., 1960, {Moving Envelopes of Stars}. Cambridge, MA: Harvard
  University Press

\bibitem[{{Stone} {et~al}\mbox{.}(2008){Stone}, {Gardiner}, {Teuben}, {Hawley},
  \& {Simon}}]{Stone2008}
{Stone} J.~M., {Gardiner} T.~A., {Teuben} P., {Hawley} J.~F., {Simon} J.~B.,
  2008, \apjs, 178, 137

\bibitem[{Toro {et~al}\mbox{.}(1994)Toro, Spruce, \& Speares}]{Toro1994}
Toro E., Spruce M., Speares W., 1994, Shock Waves, 4, 25

\bibitem[{{{\v S}urlan} {et~al}\mbox{.}(2012){{\v S}urlan}, {Hamann},
  {Kub{\'a}t}, {Oskinova}, \& {Feldmeier}}]{Surlan2012}
{{\v S}urlan} B., {Hamann} W.-R., {Kub{\'a}t} J., {Oskinova} L.~M., {Feldmeier}
  A., 2012, \aap, 541, A37

\bibitem[{Vink(2015)}]{Vink2015}
Vink J.~S., 2015, in Astrophysics and Space Science Library, Vol. 412, Very
  Massive Stars in the Local Universe, Vink J.~S., ed., Springer International
  Publishing, pp. 77--111

\bibitem[{{Vink} {et~al}\mbox{.}(1999){Vink}, {de Koter}, \&
  {Lamers}}]{Vink1999}
{Vink} J.~S., {de Koter} A., {Lamers} H.~J.~G.~L.~M., 1999, \aap, 350, 181

\bibitem[{{Vink} {et~al}\mbox{.}(2000){Vink}, {de Koter}, \&
  {Lamers}}]{Vink2000}
{Vink} J.~S., {de Koter} A., {Lamers} H.~J.~G.~L.~M., 2000, \aap, 362, 295

\end{thebibliography}
